\documentclass[12pt]{article}\usepackage{QC_Inf}

\begin{document}

\title{Quantum Computer as a\\
Probabilistic Inference Engine}

\author{Robert R. Tucci\\
        P.O. Box 226\\
        Bedford,  MA   01730\\
        tucci@ar-tiste.com}

\date{ \today}

\maketitle

\vskip2cm
\section*{Abstract}
We propose a new class of quantum computing
algorithms which
 generalize many standard
ones.
The goal of our algorithms is to estimate
probability distributions.
Such estimates are useful in, for example,
applications of
Decision Theory and Artificial Intelligence,
where inferences are made based on uncertain
knowledge.
The class of algorithms that we propose
is based on a construction method that
generalizes
a Fredkin-Toffoli (F-T) construction method used
in the field of classical reversible computing.
F-T showed how, given any binary deterministic
circuit, one can construct another binary
deterministic
circuit which does the same calculations in a
reversible manner. We show how, given any
classical stochastic network (classical Bayesian
net), one can construct a quantum network (quantum
Bayesian net).
By running this quantum Bayesian net
on a quantum computer, one can
calculate any conditional probability
that one would be interested
in calculating for the original
classical Bayesian net.
Thus,
we generalize
the F-T construction method so
that it can be applied to any
classical stochastic circuit, not
just binary deterministic ones.
We also show that, in certain situations,
 our class of algorithms
can be combined with Grover's algorithm
to great advantage.

\newpage
\section{Introduction}
In this paper, we use the language
of classical Bayesian (CB) and
quantum Bayesian (QB) nets\cite{Tucci-review}.
The reader is expected to possess
a rudimentary command of this language.

We begin this paper with a review of
various standard quantum computing algorithms;
 namely, those due to
Deutsch-Jozsa\cite{D-J},
 Simon\cite{Simon},
 Bernstein-Vazirani\cite{B-V},
and Grover\cite{Grover}. We discuss these
standard algorithms both in terms of
qubit circuits (the conventional approach) and
QB nets.
Then we propose a class of quantum computing
algorithms which
 generalizes the standard ones.

 Most standard quantum computing algorithms are
 designed for calculating
 deterministic or almost deterministic
probability  distributions.
(By a deterministic probability
distribution we mean one whose range is
restricted to either zero or unit probabilities.)
In contrast, our algorithms can also
estimate more general probability
distributions. Such estimates are
useful in, for example,
applications of Decision Theory and
 Artificial Intelligence, where
inferences are made based on uncertain knowledge.

Since some of the standard
 algorithms are contained in the class of
algorithms that we propose,
some algorithms in our class
have a time-complexity advantage over  the
best classical algorithms for
performing the same task.
Even those
algorithms in our class that
have no complexity advantage might still be useful
for nanoscale quantum computing because they are
 reversible and thus dissipate less power. Power
dissipation is best minimized in nanoscale
 devices since it can lead
to serious performance degradation.

The class of algorithms that we propose
 in this paper is based on
a construction method that
generalizes a Fredkin-Toffoli (F-T)
construction method\cite{F-T} used
 in the field of classical
reversible computing.
F-T showed in Refs.\cite{F-T} how,  given any
binary gate $f$ (i.e., a function
$f:\{0,1\}^r\rarrow \{0,1\}^s$,
for some integers $r,s$),
one can construct another binary
gate $\overline{f}$ such that $\overline{f}$
 is a deterministic reversible extension (DRE)
of $f$. $\overline{f}$ can be used
to perform the same
calculations as $f$, but in a
reversible manner. Binary gates $f$ and
$\overline{f}$ can
 be represented as binary
deterministic circuits. In this paper,
 we show how,  given
 any CB net $\cbnet$, one can construct a QB net
 $\qbnet$ which is a ``q-embedding" (q=quantum) of
$\cbnet$. By running $\qbnet$
on a quantum computer, one can calculate
any conditional probability
that one would be interested
in calculating for the CB net $\cbnet$.
Our method for constructing
a q-embedding for a
CB net is a generalization of
the  F-T method for constructing a
DRE of a binary deterministic
circuit. Thus, we generalize their method
so that it applies to any classical
stochastic circuit, not just
binary deterministic ones.

A quantum compiler
\cite{Barenco}
\cite{Tucci-qubiter}
can ``compile" a unitary
matrix; i.e., it can express the matrix
as a SEO (sequence of
elementary operations) that a quantum computer
can understand. To run a QB net on a quantum computer,
we need to replace the QB net
 by an equivalent SEO\cite{Tucci-how-to-compile}.
 This can be done with the help
of a quantum compiler.
Thus,
the class of algorithms that we propose
promises to be fertile ground
for the use of quantum compilers.

In certain cases,
the probabilities that we
wish to find are too small
to be measurable by
running $\qbnet$ on a quantum
computer. However,
we will show that sometimes
 it is possible to
define a new QB net, call it
$\qbnet'$, that magnifies
and makes measurable the
probabilities that were
unmeasurable using $\qbnet$ alone.
We will refer to $\qbnet'$ as
Grover's
Microscope for $\qbnet$, because
$\qbnet'$ is closely
related to Grover's algorithm,
and it magnifies the
probabilities found with $\qbnet$.

\section{Notation and Other
Preliminaries}\label{sec:notation}

In this section, we will introduce certain
notation that is
used throughout the paper.

We will use the word ``ditto"
as follows. If we say ``A (ditto, X) is smaller than B (ditto,
Y)", we mean
``A is smaller than B" and ``X is smaller than Y".

Let $Bool = \{0,1\}$.
For integers $a$ and $b$ such that $a\leq b$,
let $Z_{a,b} = \{a, a+1, a+2, \ldots b\}$.

For any statement ${\cal S}$,
we define the truth function $\theta({\cal S})$
to equal 1 if ${\cal S}$ is true
and 0 if ${\cal S}$ is false.
For example, $\theta(x>0)$ represents the unit step
function and $\delta(x, y)=\theta(x=y)$
the Kronecker delta function.

$\oplus$ will denote
addition mod 2. For any integer $x$,
$x\%2$ will mean the remainder from
dividing $x$ by 2.  For example,
$4\%2=0$ and $5\%2=1$. (This
same $\%$ notation is used
in the C programming language.)
When speaking of bits with states
0 and 1, we will often use an overbar to represent
the opposite state:
$\bar{0} = 1$, $\bar{1} = 0$.
Note that if $x, k\in Bool$ then

\beq
\sum_k (-1)^{kx} = 1 + (-1)^x = 2 \delta(x, 0)
\;.
\label{eq:binary-orthogonality}\eeq
If $\vec{x},\vec{y}\in Bool^n$,
we will use $\vec{x}\cdot \vec{y}=
 \sum_{\alpha=0}^{n-1} x_\alpha y_\alpha$,
 where the addition is normal, not mod 2.

Given $x\in Z_{0,\infty}$,
let $x= \sum_{\alpha=0}^{\infty} x_\alpha 2^\alpha$,
where $x_\alpha \in Bool$
for all $\alpha$.
Then we will denote
 the binary representation of $x$ by
$bin(x) = (x_0, x_1, x_2, \ldots)$.
Thus, $bin_\alpha(x) = x_\alpha$.

On the other hand, given
$\vec{x}=(x_0, x_1, x_2, \ldots)
\in Bool^\infty$,
let
$x= \sum_{\alpha=0}^{\infty} x_\alpha 2^\alpha$.
Then we will denote
 the decimal representation of $\vec{x}$ by
 $dec(\vec{x})=x$.

We will
use the symbol
$\sum_{\cdot}$ to denote a sum of
whatever is on the right hand side of this symbol
over those indices  with a dot underneath them.
For example,
$\sum_\cdot f(\punto{a}, \punto{b}, c)=
\sum_{a,b} f(a, b, c)$.
Furthermore, $\sum_{all}$ will denote a
sum over all indices.
If we wish to exclude a particular index from the
summation,
we will indicate this by a slash followed by the
name of the index.
For example, in $\sum_{all/a,b}$ we
wish to exclude summation over $a$ and $b$.
Suppose $f$ maps set $S$ into the
 complex numbers. We
will often use
$\frac{f(x)}
{\sum_x num}$
to represent $\frac{f(x)}
{\sum_{x\in S} f(x)}$. Thus, $num$ is shorthand
for the numerator of the fraction.

We will underline random variables.
$P(\rva = a)=P_\rva(a)$ will denote
 the probability
that
the random variable $\rva$ assumes value $a$.
$P(\rva=a)$ will often be abbreviated by $P(a)$
when no
confusion will arise. $S_\rva$ will denote the
set of values which the random variable $\rva$
may assume, and $N_\rva$ will denote the number of
elements in $S_\rva$. $pd(B|A)$ will stand
for the set of probability
distributions $P(\cdot|\cdot)$ such that
$P(b|a)\geq 0$ and
 $\sum_{b' \in B}P(b'|a)=1$
 for all $a\in A$ and $b\in B$.

This paper will also utilize certain notation
and nomenclature
associated with classical and quantum Bayesian
nets. For example, we will use $(x_\cdot)_A$
to denote $\{x_i : i\in A \}$.
See Ref.\cite{Tucci-review} for a review of
such notation.

$H_1 = \left (\begin{array}{rr} 1 & 1\\ 1 & -1
\end{array} \right) $
 is the one bit Hadamard matrix.
 $H_\nb = H_1^{\otimes \nb}$ (the
 n-fold tensor product of $H_1$)
 is the $\nb$ bit Hadamard matrix.
 We will also use
 $\hat{H}_1= \frac{1}{\sqrt{2}}H_1$ and
 $\hat{H}_\nb = \hat{H}_1^{\otimes \nb} =
 \frac{1}{\sqrt{2^\nb}}H_1^{\otimes \nb}$.
 Note that $(H_1)_{b,b'} = (-1)^{bb'}$
 for $b, b'\in Bool$,
 and $(H_\nb)_{\vec{b},\vec{b'}} =
 (-1)^{\vec{b}\cdot \vec{b'}}$
 for $\vec{b}, \vec{b'}\in Bool^\nb$.

Any $2 \times 2$ matrix $M$ which acts on bit
 $\alpha$ will be denoted by $M(\alpha)$.
(We like to use lower case Greek letters
for bit labels.)
In this notation, a controlled-not (cnot) gate
with control bit $\kappa$ and
target bit $\tau$ can be expressed as
$\sigma_x(\tau)^{\nop(\kappa)}$. See
Ref.\cite{Tucci-qubiter} for more details about
this notation.

 Let $\vec{\kappa}=(\kappa_0, \kappa_1, \ldots,
 \kappa_{\nb-1})$
label  $\nb$ bits.
Assume all $\kappa_i$ are distinct.
We will often use $\ns=2^\nb$,
where $\nb$ stands for number of bits and
$\ns$ for number of states.
If $ \ket{\phi}_{\kappa_i}=\ket{\phi(\kappa_i)}$
is a ket for qubit $\kappa_i$, define
$\ket{\phi}_{\vec{\kappa}}=\ket{\phi(\vec{\kappa})
} =
\prod_{i=0}^{\nb-1}\ket{\phi(\kappa_i)}$.
For example, if

\beq
\ket{0}_{\kappa_i}=
\left(\begin{array}{c}1\\0\end{array}\right)
\;
\eeq
 for all $i$, then

\beq
\ket{0}_{\vec{\kappa}} =
\prod_{i=0}^{\nb-1}\ket{0}_{\kappa_i}=
\left(\begin{array}{c}1\\0\end{array}\right)
\otimes
\left(\begin{array}{c}1\\0\end{array}\right)
\otimes\cdots\otimes
\left(\begin{array}{c}1\\0\end{array}\right)
=
[1,0,0,\ldots,0]^T
\;.
\eeq
Likewise, if $\Omega(\kappa_i)$ is an operator
acting on qubit
$\kappa_i$,
define $\Omega(\vec{\kappa}) = \prod_{i=0}^{\nb-1}
\Omega(\kappa_i)$. For example,
$H_1(\vec{\kappa})= \prod_{i=0}^{\nb-1} H_1(\kappa_i)$
is an $\nb$ bit Hadamard matrix.

 Next, we will introduce some notation
 related to  Pauli matrices.
 The Pauli matrices are given by:

\beq
\sigma_x =
\left(
\begin{array}{cc}
0&1\\
1&0
\end{array}
\right)
\;,\;\;
\sigma_y =
\left(
\begin{array}{cc}
0&-i\\
i&0
\end{array}
\right)\;,\;\;
\sigma_z =
\left(
\begin{array}{cc}
1&0\\
0&-1
\end{array}
\right)
\;.
\eeq
If $\ket{+_z}$ and $\ket{-_z}$ represent
the eigenvectors of
$\sigma_z$ with
eigenvalues $+1$ and $-1$, respectively,
then we define

\beq
\ket{0} =\ket{+_z} =
\left(
\begin{array}{c}
1\\0
\end{array}
\right)
\;,
\eeq
and

\beq
\ket{1} =\ket{-_z} =
\left(
\begin{array}{c}
0\\1
\end{array}
\right)
\;.
\eeq
We denote the ``number operator" by $\nop$. Thus

\beq
\nop =  \left(\begin{array}{cc}
0&0\\
0&1
\end{array}\right)
=\ket{-_z}\bra{-_z}=  \frac{1-\sigma_z}{2}
\;,
\eeq
and

\beq
\nbar = 1-\nop = \left(\begin{array}{cc}
1&0\\
0&0
\end{array}\right)
=\ket{+_z}\bra{+_z}=  \frac{1+\sigma_z}{2}
\;.
\eeq
Since $n$ and $\sigma_z$ are diagonal, it is
easy to see that

\beq
(-1)^\nop = \sigma_z
\;.
\eeq
It is also useful to introduce symbols
for the projectors with respect to $\ket{0}$
and $\ket{1}$;

\beq
P^z_0 = \ket{0}\bra{0} = \nbar
\;,
\eeq

\beq
P^z_1 = \ket{1}\bra{1} = \nop
\;.
\eeq

Most of the definitions and results stated so far
for $\sigma_z$
have counterparts for
$\sigma_x$ and $\sigma_y$.
The counterpart results can be
easily proven by applying a rotation that
interchanges the coordinate axes.
Let $w\in \{x, y, z\}$.
If $\ket{+_w}$ and $\ket{-_w}$
represent the eigenvectors of $\sigma_w$ with
eigenvalues $+1$ and $-1$, respectively,
then we define

\beq
\ket{0_w} =\ket{+_w}
\;,
\eeq
and

\beq
\ket{1_w} =\ket{-_w}
\;.
\eeq
Let

\beq
\nop_w=
\ket{-_w}\bra{-_w}=  \frac{1-\sigma_w}{2}
\;,
\eeq

\beq
\nbar_w = 1-\nop_w
=\ket{+_w}\bra{+_w}=  \frac{1+\sigma_w}{2}
\;.
\eeq
As when $w=z$, one has

\beq
(-1)^{\nop_w} = \sigma_w
\;.
\eeq
Let

\beq
P^w_0 = \ket{0_w}\bra{0_w} = \nbar_w
\;,
\eeq
and

\beq
P^w_1 = \ket{1_w}\bra{1_w} = \nop_w
\;.
\eeq

\begin{figure}[h]
    \begin{center}
    \epsfig{file=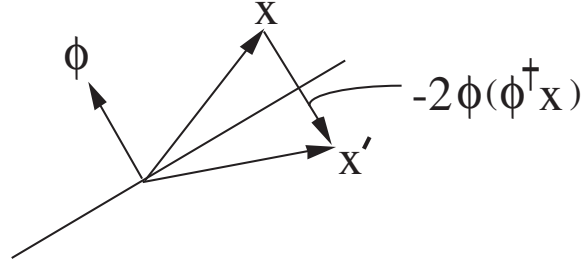, height=1.5in}
    \caption{Reflection with respect to plane
    perpendicular to $\phi$.}
    \label{fig:reflection}
    \end{center}
\end{figure}

In understanding Grover's algorithm,
 it is helpful
 to be aware of
some simple properties of
reflections on a plane.
Suppose $\phi$ is a normalized ($\phi^\dagger\phi=1$)
complex vector. Define
the projection and
reflection operators for $\phi$ by

\beq
\Pi_\phi = \phi\phi^\dagger\;,\;\;
R_\phi= 1 - 2\Pi_\phi
\;.
\eeq
Note that $\Pi_\phi^2 = \Pi_\phi$.
Fig.\ref{fig:reflection} shows that if
$x' = R_\phi x$, then
$x'$ is the reflection of $x$ with
respect to the  plane
perpendicular to $\phi$. For example,
 $R_\phi \phi = -\phi$.

Some simple properties of $R_\phi$ are
as follows.
$R_\phi = R_\phi^\dagger$ and
$R_\phi R_\phi^\dagger = R_\phi^2 = 1$.
Since reflections are unitary matrices,
a product of reflections is also a unitary matrix.

Note that

\begin{subequations}
\begin{eqnarray}
(-1)^{\Pi_\phi} &=& e^{i \pi \Pi_\phi}
= 1 + (e^{i \pi \Pi_\phi}-1) \\
&= &1 + \Pi_\phi(e^{i \pi }-1) =1 - 2 \Pi_\phi \label{eq:proj-taylor}\\
&=&R_\phi
\;.
\end{eqnarray}
\end{subequations}
(Eq.(\ref{eq:proj-taylor})
follows from the Taylor expansion of $e^{i \pi \Pi_\phi}$.)

If $e_1, e_2, \ldots, e_n$ is an orthonormal
basis for a
vector space,
$\Pi_i = e_i e_i^\dagger$, and
$R_i = 1-2\Pi_i$, then
the product of the $R_i$
in any order  is $-1$. Indeed,

\begin{subequations}
\begin{eqnarray}
R_1 R_2\ldots R_n &=&
(1 - 2\Pi_1)(1-2\Pi_2)\ldots (1-2\Pi_n)\\
&=& 1 - 2(\Pi_1 + \Pi_2 +\ldots \Pi_n)\\
&=& -1
\;.
\end{eqnarray}
\end{subequations}

Another property of reflection operators which
is useful for understanding Grover's algorithm
is the following. Let

\beq
e_0=
\left(
\begin{array}{c}
1\\0
\end{array}
\right),
\;\;
e_1=
\left(
\begin{array}{c}
0\\1
\end{array}
\right)
\;.
\eeq
Now suppose that $e'_1$
is obtained by rotating
$e_1$ clockwise
 by an angle $\theta/2$:

\beq
e'_1=
\left(
\begin{array}{cc}
\cos (\theta/2) & \sin (\theta/2)\\
-\sin(\theta/2) & \cos(\theta/2)
\end{array}
\right) e_1
=
\left(
\begin{array}{c}
\sin (\theta/2)\\
\cos (\theta/2)
\end{array}
\right)
\;.
\eeq
$e_1'\approx e_1$ for small $\theta$.
It is easy to check that
the double reflection
$-R_{e'_1} R_{e_0}$ is equivalent to a rotation
(also clockwise) by $\theta$:

\beq
-R_{e'_1} R_{e_0} =
\left(
\begin{array}{cc}
\cos \theta & \sin \theta\\
-\sin \theta & \cos \theta
\end{array}
\right)
\;.
\label{eq:double-reflection-gen}
\eeq
(That these two successive
reflections equal a rotation
was to be expected, since the reflections
are orthogonal matrices and a product
of orthogonal matrices is itself orthogonal.)

Above, we have considered plane reflections $R_\phi$
acting on a complex vector space, but our formulas
still hold true when $R_\phi$ acts on a real instead
of a complex vector space.
In the case of real vector spaces,
the
Hermitian conjugate symbol $\dagger$
is replaced by
 the matrix transpose symbol $T$,
and unitary matrices are replaced by orthogonal matrices.

\section{Some Standard Algorithms}
Next we will discuss several standard algorithms
that are considered among the best that the
quantum computation field has to offer at the
present time. Later,
we will try to generalize these standard
algorithms.
\subsection{Deutsch-Jozsa Algorithm}

In this section we will discuss the D-J
(Deutsch-Jozsa)
 algorithm\cite{D-J}.
We will do this first in terms of
qubit circuits (the conventional approach),
and then in terms of QB nets.

\begin{figure}[h]
    \begin{center}
    \epsfig{file=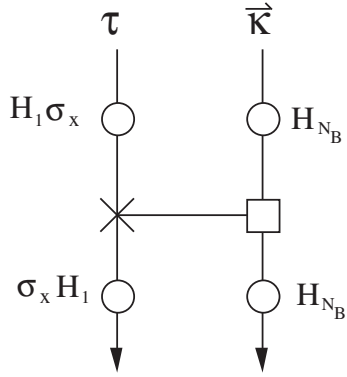, height=2in}
    \caption{Qubit circuit for D-J's algorithm.}
    \label{fig:D-J-circuit}
    \end{center}
\end{figure}

Let $\vec{\kappa}=(\kappa_0, \kappa_1, \ldots,
 \kappa_{\nb-1})$
label  $\nb$ ``control" bits
and let $\tau$ label  a
 single ``target" bit.
 Assume that $\tau$ and
 all the $\kappa_i$ are distinct.
We will denote the state
of these bits in the preferred basis
(the eigenvectors of $\sigma_z$)
by $\ket{x}_{\vec{\kappa}} \ket{y}_\tau$,
where $x \in Bool^\nb$ and $y\in Bool$.
Given a function $f:Bool^\nb\rarrow Bool$,
define the unitary operator $\Omega$ by

\beq
\Omega=
\sx(\tau)\hat{H}_1(\tau)\hat{H}_1(\vec{\kappa})
\sx^{f(\vec{\nop}(\vec{\kappa}))}(\tau)
\hat{H}_1(\vec{\kappa})\hat{H}_1(\tau)\sx(\tau)
\;,
\label{eq:D-J-omega-def}\eeq
where $\vec{\nop}(\vec{\kappa}) = (\nop(\kappa_0),
\nop(\kappa_1), \ldots, \nop(\kappa_{\nb-1}))$.
 The operation
$\sx^{f(\vec{\nop}(\vec{\kappa}))}(\tau)$, because
it
depends on $f$,
is often called an ``oracle"
and each use of it is called a ``query".
The right hand side of Eq.(\ref{eq:D-J-omega-def})
may be represented by
the circuit diagram
shown in Fig.\ref{fig:D-J-circuit}.
The D-J algorithm consists of applying $\Omega$
to an initial state $\ket{0}_{\vec{\kappa}}
\ket{0}_\tau$ of bits $\vec{\kappa}$ and $\tau$,
and then measuring
the final state of these bits in the preferred
basis.

Fig.\ref{fig:D-J-circuit} and the right hand side
of
Eq.(\ref{eq:D-J-omega-def}) are two equivalent ways
of
representing a particular SEO.
There are infinitely many SEOs that yield
$\Omega$.
Fig.\ref{fig:D-J-circuit} is just one of them.
In fact, the original D-J paper\cite{D-J} gave a
different SEO for $\Omega$, one with
two queries instead of one.

For $X \in Bool^\nb, Y\in Bool$, let

\beq
\ket{\psi_0} = \ket{X}_{\vec{\kappa}} \ket{Y}_\tau
\;,
\eeq
and

\beq
\ket{\psi_i} = \Omega_i \ket{\psi_{i-1}}
\;\;{\rm for} \;\;
i=1, 2, \ldots
\;,
\eeq
where

\beq
\Omega_1 =
\hat{H}_1(\vec{\kappa})\hat{H}_1(\tau)\sx(\tau)
\;,
\eeq

\beq
\Omega_2 = \sx^{f(\vec{\nop}(\vec{\kappa}))}(\tau)
\;,
\eeq
and

\beq
\Omega_3 =  \Omega_1^\dagger
\;.
\eeq
Then it is easy to show using simple
identities (such as $(H_1)_{b,b'}= (-1)^{bb'}$,
$\overline{0} = 1$, $\overline{1} = 0$, and
$(-1)^b = (-1)^{-b}$ for $b, b'\in Bool$)
 that

\beq
\ket{\psi_1} = \frac{1}{\sqrt{2^{\nb + 1}}}
 \sum_{x, y } (-1)^{x\cdot  X + y\overline{Y}}
 \ket{x}_{\vec{\kappa}}\ket{y}_\tau
\;,
\eeq

\beq
\ket{\psi_2} = \frac{1}{\sqrt{2^{\nb + 1}}}
\sum_{x, y } (-1)^{x\cdot  X + y\overline{Y}}
\ket{x}_{\vec{\kappa}}\ket{y \oplus f(x)}_\tau
\;,
\eeq

\beq
\ket{\psi_3} = \frac{1}{2^{\nb + 1}} \sum_{x, y,
X', Y' }
(-1)^{x\cdot (X'- X)+ y(\overline{Y'}
 - \overline{Y}) + \overline{Y'} f(x)    }
\ket{X'}_{\vec{\kappa}}\ket{Y'}_\tau
\;.
\label{eq:psi3}
\eeq
Applying $\bra{X',Y'}$ to the right hand side
of Eq.(\ref{eq:psi3}) and using the identity
Eq.(\ref{eq:binary-orthogonality}) finally yields:

\beq
\av{ X', Y' | \Omega | X, Y} =
\delta(Y', Y) \frac{1}{2^\nb}
\sum_{x\in Bool^\nb} (-1)^{x\cdot  (X'- X) +
\overline{Y'} f(x)    }
\;
\label{eq:D-J-omega-mat-elem}\eeq
for all $X', X\in Bool^\nb$ and   $Y', Y\in
Bool$.
Thus,
if the initial states
of $\vec{\kappa}$ and $\tau$ are $\rvX = 0$  and
$\rvY = 0$,
then the probability of obtaining $\rvX'= X'$
for the final state of $\vec{\kappa}$  is

\beqa
P(X'| \rvX = \rvY = 0) &=&
\sum_{Y'}
| \av{ X', Y' | \Omega | X=0, Y=0} |^2 \nonumber\\
&=&
\frac{1}{4^\nb}
| \sum_{x} (-1)^{x\cdot  X'+ f(x)} |^2
\;.
\label{eq:D-J-prob}
\eeqa

Let ${\cal F}_{bal}$, the set of ``balanced"
functions,
be the set of all $f:Bool^\nb \rarrow Bool$
such that $f$
maps exactly half of its domain to zero and half
to one.
Let ${\cal F}_{con}$, the set of ``constant"
functions,
be the set of all $f:Bool^\nb \rarrow Bool$
such that $f$ maps all its domain to zero or all
of it to one.
From Eq.(\ref{eq:D-J-prob}),
if $X'=0$ and $f\in {\cal F}_{bal} \cup {\cal
F}_{con}$, then

\beq
P(\rvX'=0| \rvX = \rvY = 0) =
\left \{
\begin{array}{ll}
1 & {\rm if}\;\;f\in{\cal F}_{con}\\
0 & {\rm if}\;\;f\in{\cal F}_{bal}
\end{array}
\right .
\;.
\label{eq:D-J-prob2}\eeq

\begin{figure}[h]
    \begin{center}
    \epsfig{file=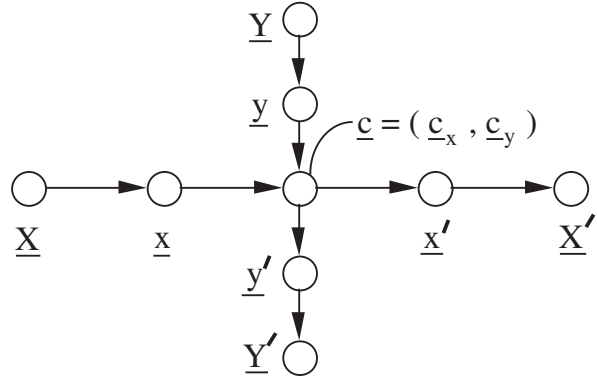, height=2in}
    \caption{QB net for D-J's algorithm.}
    \label{fig:D-J-qbnet}
    \end{center}
\end{figure}

Now consider the QB net defined by
 Fig.\ref{fig:D-J-qbnet}
and Table
\ref{tab:D-J-qbnet}.

\BeginQBNetTabular      \label{tab:D-J-qbnet}
    $\rvX$ & $X \in Bool^\nb$ & $\delta(X,0)$ & \\
\hline
    $\rvY$ & $Y \in Bool$ & $\delta(Y,0)$ & \\
\hline
    $\rvx$ & $x\in Bool^\nb$ & $(-1)^{x\cdot
X}/{\sqrt{2^\nb}}$ & \\
\hline
    $\rvy$ & $y\in Bool$ & $(-1)^{y
\overline{Y}}/{\sqrt{2}}$ & \\
\hline
    $\rvc$ & $c=(c_x, c_y), c_x \in Bool^\nb, c_y
\in Bool$ & $\delta(c_x, x)\delta(c_y, y
\oplus f(x))$ & \\
\hline
    $\rvx'$ & $x'\in Bool^\nb$ & $\delta(x', c_x)$
& \\
\hline
    $\rvy'$ & $y'\in Bool$ & $\delta(y', c_y)$ &
\\
\hline
    $\rvX'$ & $X'\in Bool^\nb$ & $(-1)^{X'\cdot
x'}/{\sqrt{2^\nb}}$ & \\
\hline
    $\rvY'$ & $Y'\in Bool$ &
$(-1)^{\overline{Y'}y'}/{\sqrt{2}}$ & \\

\EndQBNetTabular

For this net, the amplitude $A(x.)$ of net story
$x.$ is
the product of all the terms in the third column
of Table \ref{tab:D-J-qbnet}.
If
 $\rvX = 0$  and $\rvY = 0$,
then the probability of obtaining $\rvX'= X'$ is

\beq
P(X'| \rvX = \rvY = 0) =
\frac{
\sum_{Y'} \left |
\sum_{all/X'Y', X, Y}
A(x.)|_{X=Y=0}
\right | ^2
}
{ \sum_{X'} num}
\;,
\label{eq:D-J-a2p}\eeq
where $A(x.)$ on the right hand side is evaluated
at $X=Y=0$.
Substituting the value of $A(x.)$ into
Eq.(\ref{eq:D-J-a2p}) immediately
yields Eq.(\ref{eq:D-J-prob}).

Note that one can calculate the
probability
distribution
Eq.(\ref{eq:D-J-prob}) by
means of a CB net
  instead of a QB net.
One can do this with the CB net defined by
the graph
$\rvX'\rarrow \rvY'$,
with:

\BeginCBNetTabular      \label{tab:D-J-cbnet}
    $\rvX'$ & $X' \in Bool^\nb$ & $P_{\rvX'}(X')$ &
 \\
\hline
    $\rvY'$ & $Y'\in Bool$ &
    $P_{\rvY'|\rvX'}(Y'|X')$ & \\
\EndQBNetTabular
where
$P_{\rvX'}$ and $P_{\rvY'|\rvX'}$
are calculated from

\beq
P_{\rvX',\rvY'}(X',Y') =
| \av{X', Y' | \Omega | X=0, Y=0 } |^2
\;.
\eeq
We will say that the CB net
defined by the graph
$\rvX'\rarrow \rvY'$ and Table
\ref{tab:D-J-cbnet}
is ``q-embedded" in the QB net defined by
Fig.\ref{fig:D-J-qbnet} and
Table \ref{tab:D-J-qbnet}.
In subsequent sections, we will say much
more about q-embedding  of CB nets.

\subsection{Simon's Algorithm}

In this section we will discuss Simon's
algorithm\cite{Simon}.
We will do this first in terms of
qubit circuits (the conventional
approach),
and then in terms of QB nets.

\begin{figure}[h]
    \begin{center}
    \epsfig{file=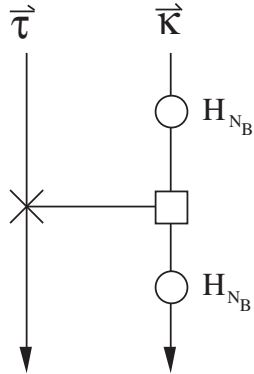, height=2in}
    \caption{Qubit circuit for Simon's algorithm.}
    \label{fig:Si-circuit}
    \end{center}
\end{figure}

Simon's algorithm uses $\nb$ ``control" bits, just
like the D-J algorithm.
However, it uses $\nb$ target bits whereas the D-J
algorithm uses only one.
Simon's algorithm deals with a vector-valued
function
 $f:Bool^\nb\rarrow Bool^\nb$, whereas D-J's
algorithm deals with a scalar-valued function
$f:Bool^\nb\rarrow Bool$.
Let $\vec{\kappa}=(\kappa_0, \kappa_1, \ldots,
 \kappa_{\nb-1})$
label  $\nb$ ``control" bits
and let $\vec{\tau}=(\tau_0, \tau_1, \ldots,
 \tau_{\nb-1})$ label  $\nb$
  ``target" bits. Assume all $\tau_i$ and
 $\kappa_i$ are distinct. We will denote the state
of these bits in the
preferred basis (the eigenvectors of $\sigma_z$)
by
$\ket{x}_{\vec{\kappa}} \ket{y}_{\vec{\tau}}$,
where $x \in Bool^\nb$ and $y\in Bool^\nb$.
Given a function $f=(f_0, f_1, \ldots, f_{\nb-1})$
where
$f_i:Bool^\nb\rarrow Bool$,
define the unitary operator $\Omega$ by

\beq
\Omega=
\hat{H}_1(\vec{\kappa})
\left [\prod_{i=0}^{\nb-1}
\sx^{f_i(\vec{\nop}(\vec{\kappa}))}(\tau_i) \right
]
\hat{H}_1(\vec{\kappa})
\;.
\label{eq:s-omega-def}\eeq
The operator $\Omega$ for Simon's algorithm
is analogous to the $\Omega$ defined by
Eq.(\ref{eq:D-J-omega-def}) for the D-J algorithm.
The right hand side of Eq.(\ref{eq:s-omega-def})
may be represented by
the circuit diagram of Fig.\ref{fig:Si-circuit}.
Simon's algorithm consists of applying
$\Omega$ given by  Eq.(\ref{eq:s-omega-def})
to an initial state $\ket{0}_{\vec{\kappa}}
\ket{0}_{\vec{\tau}}$ of bits $\vec{\kappa}$
and $\vec{\tau}$, and then measuring
the final state of these bits in the preferred
basis. One performs this routine several
times. The measurement outcomes  allow one to
determine
the period of the function $f$ if  $f$ is of a
special periodic type
that will be specified later.

Using the same techniques that we used to evaluate
the matrix elements of $\Omega$ for the D-J
algorithm,
one finds

\beq
\av{ X', Y' | \Omega | X, Y} =
\frac{1}{2^\nb}
\sum_{x\in Bool^\nb} (-1)^{x\cdot  (X'- X)  }
\delta(Y' , Y \oplus f(x))
\;,
\label{eq:s-omega-mat-elem}\eeq
for all $ X', Y', X, Y\in Bool^\nb$.
If the initial states
of $\vec{\kappa}$ and $\vec{\tau}$ are $\rvX = 0$
 and $\rvY = 0$,
then the probability of obtaining $\rvX'= X'$
for the final state of $\vec{\kappa}$  is

\beqa
P(X'| \rvX = \rvY = 0) &=&
\sum_{Y'}
| \av{ X', Y' | \Omega | X=0, Y=0} |^2 \nonumber\\
&=&\frac{1}{4^\nb}
\sum_{Y'}
| \sum_{x} (-1)^{x\cdot  X'}
\delta(Y', f(x))
|^2
\;.
\label{eq:s-prob}\eeqa

Now suppose ${\cal F}_{S}$ is the set of those
functions
$f:Bool^\nb \rarrow Bool^\nb$ such that $f$
is 2 to 1 (i.e., $f$ maps
exactly two domain points
into each image point) and has a ``period"
$\Delta$. By a period
$\Delta$, we mean a non-zero element
of $Bool^\nb$ such that
$f(x) = f(x \oplus \Delta)$ for all $x\in
Bool^\nb$.
For any $f \in {\cal F}_{S}$ and any
$y\in Bool^\nb$, there exist exactly two elements
of $Bool^\nb$,
call them $x_1$ and $x_2$, such that $x_1 = x_2
\oplus \Delta$ and
$f(x_1) = f(x_2) = y$.
Call $f^{-1}_p(y)$ one of these
$x$ values, and call
$f^{-1}_p(y) \oplus \Delta$ the other.
(The $p$ subscript stands for ``principal part",
in analogy with Complex Analysis.)
If $f \in {\cal F}_{S}$,
and $I(f)$ is the image of $f$, then

\beq
\delta(Y', f(x)) =
\left \{
\begin{array}{ll}
\delta(f^{-1}_p(Y'), x)
+
\delta(f^{-1}_p(Y') \oplus \Delta, x) ,
& {\rm if} \;\; Y' \in I(f) \\
0 &{\rm otherwise}
\end{array}
\right.
\;.
\eeq
Substituting this expression for
$\delta(Y', f(x))$
into
Eq.(\ref{eq:s-prob}) and using
Eq.(\ref{eq:binary-orthogonality})
yields

\beq
P(X'| \rvX = \rvY = 0) =
\frac{1}{2^{\nb-1}}
\delta(X'\cdot  \Delta, 0)
\;.
\label{eq:s-prob2}\eeq
To calculate the period $\Delta$ of $f$,
run the experiment $\nu$ times,
measuring $X'$ each time.
Let $X'(i)$ represent the $i$th measurement
outcome.
Then, for sufficiently large $\nu$, one can find
$\Delta$
by solving the equations $X'(1)\cdot  \Delta = 0$,
$X'(2)\cdot  \Delta = 0$, ... , $X'(\nu)\cdot
\Delta = 0$.

\begin{figure}[h]
    \begin{center}
    \epsfig{file=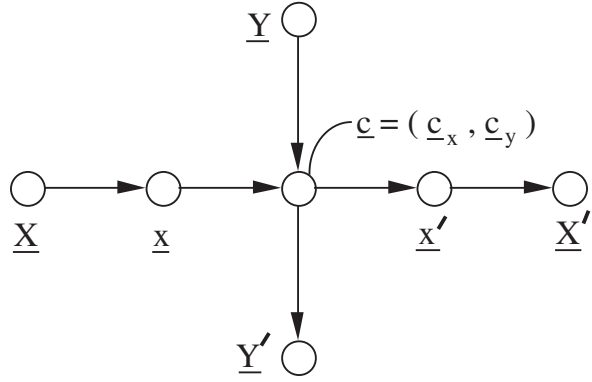, height=2in}
    \caption{QB net for Simon's algorithm.}
    \label{fig:Si-qbnet}
    \end{center}
\end{figure}

Now consider the QB net
defined by Fig.\ref{fig:Si-qbnet}
and Table \ref{tab:Si-qbnet}.

\BeginQBNetTabular      \label{tab:Si-qbnet}
    $\rvX$ & $X \in Bool^\nb$ & $\delta(X,0)$ &\\
\hline
    $\rvY$ & $Y \in Bool^\nb$ & $\delta(Y,0)$ &\\
\hline
    $\rvx$ & $x\in Bool^\nb$ & $(-1)^{x\cdot
X}/{\sqrt{2^\nb}}$ & \\
\hline
    $\rvc$ & $c=(c_x, c_y); c_x , c_y\in Bool^\nb$
& $\delta(c_x, x)\delta(c_y, Y \oplus f(x))$ &
\\
\hline
    $\rvx'$ & $x'\in Bool^\nb$ & $\delta(x', c_x)$
& \\
\hline
    $\rvX'$ & $X'\in Bool^\nb$ & $(-1)^{X'\cdot
x'}/{\sqrt{2^\nb}}$ & \\
\hline
    $\rvY'$ & $Y'\in Bool^\nb$ & $\delta(Y', c_y)$
& \\

\EndQBNetTabular

For this net, the amplitude $A(x.)$ of net story
$x.$ is
the product of all the terms in the third column
of Table \ref{tab:Si-qbnet}.
If
 $\rvX = 0$  and $\rvY = 0$,
then the probability of obtaining $\rvX'= X'$ is

\beq
P(X'| \rvX = \rvY = 0) =
\frac{
\sum_{Y'} \left |
\sum_{all/X'Y', X, Y}
A(x.)|_{X=Y=0}
\right | ^2
}
{\sum_{X'} num}
\;,
\label{eq:s-a2p}\eeq
where $A(x.)$ on the right hand side is evaluated
at $X=Y=0$.
Substituting the value of $A(x.)$ into
Eq.(\ref{eq:s-a2p}) immediately
yields Eq.(\ref{eq:s-prob}).

It is possible to calculate the
probability distribution
Eq.(\ref{eq:s-prob}) by
means of a CB net  instead of a QB net.
One can do this with the CB net
defined by
the graph
$\rvX'\rarrow \rvY'$,
with:

\BeginCBNetTabular      \label{tab:Si-cbnet}
    $\rvX'$ & $X' \in Bool^\nb$ & $P_{\rvX'}(X')$ &
 \\
\hline
    $\rvY'$ & $Y'\in Bool^\nb$ &
    $P_{\rvY'|\rvX'}(Y'|X')$ & \\
\EndQBNetTabular
where
$P_{\rvX'}$ and $P_{\rvY'|\rvX'}$
are calculated from

\beq
P_{\rvX',\rvY'}(X',Y') =
| \av{X', Y' | \Omega | X=0, Y=0 } |^2
\;.
\eeq
We will say that the CB net
defined by the graph
$\rvX'\rarrow \rvY'$ and Table
\ref{tab:Si-cbnet}
is ``q-embedded" in the QB net defined by
Fig.\ref{fig:Si-qbnet} and Table \ref{tab:Si-qbnet}.

\subsection{Bernstein-Vazirani Algorithm}

In this section we will discuss the B-V
(Bernstein-Vazirani)
 algorithm\cite{B-V}.

 To understand the B-V algorithm, it is helpful
to first establish the following simple
single qubit identities. First note that
the single qubit Hadamard matrix rotates
the $Z$-direction number operator
into the $X$-direction number operator:

\beq
\hat{H}_1 \nop_z \hat{H}_1 =
\frac{1}{2}
\left(
\begin{array}{cc}
1&-1\\
-1&1
\end{array}
\right)=
 \nop_x
\;.
\label{eq:hada-rot}
\eeq
Thus,
\beq
\sigma_x^b = [(-1)^{\nop_x}]^b = (-1)^{b \nop_x}
 =  \hat{H}_1 (-1)^{b \nop_z} \hat{H}_1
\;.
\label{eq:obf-sig-b}
\eeq
Next note that
$\sigma_x$ exchanges the components of
any vector it acts on:

\beq
\sigma_x
\left( \begin{array}{c} \alpha\\ \beta \end{array}
\right)
=
\left( \begin{array}{c} \beta\\ \alpha \end{array}
\right)
\;,
\eeq
for any complex numbers $\alpha, \beta$.
In particular,
if $b\in Bool$, then
\beq
\sigma_x^b \ket{0} = \ket{b}
\;.
\label{eq:sigma-z-flip}
\eeq

Now we are ready to discuss the B-V algorithm.
Let $\vec{\kappa}=(\kappa_0, \kappa_1, \ldots,
 \kappa_{\nb-1})$
label  $\nb$ ``control" bits
and let $\tau$ label  a
 single ``target" bit. Assume that $\tau$ and
 all the $\kappa_i$ are distinct.
We will denote the state
of these bits in the preferred basis
(the eigenvectors of $\sigma_z$)
by $\ket{x}_{\vec{\kappa}} \ket{y}_\tau$,
where $x \in Bool^\nb$ and $y\in Bool$.
For $\vec{b}\in Bool^\nb$,
define the unitary operator

\beq
\omega_{\vec{b}} =  \prod_{i=0}^{\nb-1}
\sigma_x(\kappa_i)^{b_i}
\;.
\label{eq:omega-vecb-lite}
\eeq
The B-V algorithm is simply the following
multi-qubit generalization of
Eq.(\ref{eq:sigma-z-flip})

\beq
\omega_{\vec{b}}\ket{0}_{\vec{\kappa}} =
\ket{\vec{b}}_{\vec{\kappa}}
\;.\label{eq:B-V-lite}
\eeq
That's all there is to B-V!

Eq.(\ref{eq:B-V-lite})
can be represented by a
qubit circuit
consisting of a single
wire for $\vec{\kappa}$,
with a single node representing
$\omega_{\vec{b}}$.
Eq.(\ref{eq:B-V-lite})
can also be represented by a QB net
defined by the graph $\rvX\rarrow\rvX'$,
with

\BeginQBNetTabular      \label{tab:B-V-qbnet}
    $\rvX$ & $X=(X_0, X_1, X_{\nb-1}) \in Bool^\nb$ & $\delta(X,0)$ & \\
\hline
    $\rvX'$ & $X'=(X'_0, X'_1, X'_{\nb-1}) \in Bool^\nb$ &
    $\prod_{i=0}^{\nb-1}\delta^{b_i}(X'_i, \overline{X_i})$ & \\
\EndQBNetTabular

We should mention that  it is
common in the literature to dress up
and obfuscate Eq.(\ref{eq:omega-vecb-lite})
as follows. By virtue of Eq.(\ref{eq:obf-sig-b}),
one can re-express $\omega_{\vec{b}}$ as

\beq
\omega_{\vec{b}} =
\prod_{i=0}^{\nb-1}
\sigma_x(\kappa_i)^{b_i}
=
\hat{H}_1(\vec{\kappa})
(-1)^{\sum_{i=0}^{\nb-1}
b_i\nop_z(\kappa_i)}
\hat{H}_1(\vec{\kappa})
\;.
\label{eq:obf-B-V1}
\eeq
Some workers ascend to an even higher peak of
obfuscation by adding a totally unnecessary
target qubit.
They define an operator, call it
$\Omega_{\vec{b}}$, obtained by
replacing the $(-1)$ in Eq.(\ref{eq:obf-B-V1})
by the operator $\sigma_x(\tau)$
acting on a target qubit  $\tau$:

\beq
\Omega_{\vec{b}}
=
\hat{H}_1(\vec{\kappa})
[\sigma_x(\tau)]^{\sum_{i=0}^{\nb-1}
b_i\nop_z(\kappa_i)}
\hat{H}_1(\vec{\kappa})
\;.
\eeq
At the beginning of the experiment,
they put the target qubit in a state
which is an eigenvector of $\sigma_x(\tau)$
with eigenvalue $-1$.
Thus, the obfuscated version of the B-V algorithm
with a target qubit
can be summarized by

\beq
\Omega_{\vec{b}}\ket{-_x}_\tau\ket{0}_{\vec{\kappa}}=
\omega_{\vec{b}}\ket{-_x}_\tau\ket{0}_{\vec{\kappa}}=
\ket{-_x}_\tau\ket{\vec{b}}_{\vec{\kappa}}
\;.
\eeq
We emphasize that for the B-V algorithm,
 the target qubit
 is a totally unnecessary affectation.

So far we have given an unconventional presentation of
the B-V algorithm. For completeness, we now give a
conventional
one. Define

\beq
\ket{\psi_0} = \ket{0}_{\vec{\kappa}}
\ket{-_x}_\tau
\;,
\eeq
and

\beq
\ket{\psi_i} = \Omega_i \ket{\psi_{i-1}} \;\;{\rm
for} \;\; i=1, 2, \ldots
\;,
\eeq
where

\beq
\Omega_1 =
\hat{H}_1(\vec{\kappa})
\;,
\eeq

\beq
\Omega_2 =
[\sigma_x(\tau)]^{\sum_{i=0}^{\nb-1}
b_i\nop_z(\kappa_i)}
\;,
\eeq
and

\beq
\Omega_3 = \Omega_1
\;.
\eeq
It follows that

\beq
\ket{\psi_1} =
\frac{1}{\sqrt{2^\nb}}
\sum_{\vec{x}\in Bool^\nb}
\ket{\vec{x}}_{\vec{\kappa}} \ket{-_x}_\tau
\;,
\eeq

\beq
\ket{\psi_2}=
\frac{1}{\sqrt{2^\nb}}
\sum_{\vec{x}}(-1)^{\vec{b}\cdot \vec{x}}
\ket{\vec{x}}_{\vec{\kappa}} \ket{-_x}_\tau
\;,
\eeq
and

\begin{subequations}
\label{eq:B-V-psi3}
\begin{eqnarray}
\ket{\psi_3}&=&
\frac{1}{\sqrt{2^\nb}}
\sum_{\vec{x}}(-1)^{\vec{b}\cdot \vec{x}}
\frac{1}{\sqrt{2^\nb}}
\sum_{\vec{y}}(-1)^{\vec{y}\cdot \vec{x}}
\ket{\vec{y}}_{\vec{\kappa}} \ket{-_x}_\tau\\
&=&
\frac{1}{2^\nb}
\sum_{\vec{x},
\vec{y}}(-1)^{(\vec{b}-\vec{y})\cdot \vec{x}}
\ket{\vec{y}}_{\vec{\kappa}} \ket{-_x}_\tau\\
&=&
\ket{\vec{b}}_{\vec{\kappa}}\ket{-_x}_\tau
\;.
\end{eqnarray}
\end{subequations}
To go from step (b) to step (c)
of Eq.(\ref{eq:B-V-psi3}), we used the orthogonality
property given by Eq.(\ref{eq:binary-orthogonality}).

\subsection{Grover's Algorithm}

In this section we will discuss
Grover's algorithm \cite{Grover}.

Let $\vec{\kappa}=(\kappa_0, \kappa_1, \ldots,
 \kappa_{\nb-1})$
label  $\nb$ bits.
Assume all $\kappa_i$ are distinct.
We begin by defining
the following $\ns$-dimensional
column vectors:

\beq
\ket{\mu}_{\vec{\kappa}} = \mu = \mu_\ns =
\frac{1}{\sqrt{\ns}}
[1,1,1,1 \ldots, 1]^T
=
\frac{1}{\sqrt{\ns}}
H_\nb [1,0,0,0, \ldots, 0]^T
\;,
\eeq

\beq
\ket{\phi}_{\vec{\kappa}}=\phi =
[0, \ldots, 0, 0, 1, 0, 0, \ldots, 0]^T
\;.
\eeq
All components of $\phi$ are zero except for
one predetermined component, located at position
$j_{targ}\in Z_{0, \ns-1}$,
which equals one. We will refer to $j_{targ}$
as the {\bf target state}
(not to be confused with a target qubit).
Note that we chose a special basis (or, equivalently,
a special matrix representation) from the start.
Note that
$\av{\phi | \mu} = \frac{1}{\sqrt{\ns}}$,
so $\mu$ and $\phi$
are nearly orthogonal for large $\ns$.
It is also convenient to define
the component-wise negation of $\phi$:

\beq
\ket{\phi^{not}}_{\vec{\kappa}} = \phi^{not} =
[1, \ldots, 1, 1, 0, 1, 1, \ldots, 1]^T
\;.
\eeq
Note that
$\phi^{not}$ is not normalized.

Define projection and reflection
operators for $\mu$ and $\phi$:

\beq
\Pi_\mu = \ket{\mu}\bra{\mu}
\;,\;\;
R_\mu = 1 -2 \Pi_\mu = (-1)^{\Pi_\mu}
\;,
\eeq
and

\beq
\Pi_\phi = \ket{\phi}\bra{\phi}
\;,\;\;
R_\phi = 1 -2 \Pi_\phi = (-1)^{\Pi_\phi}
\;.
\eeq
Grover's algorithm can be summarized by the following
equation\cite{Noble}\cite{QR}:

\beq (-R_\mu R_\phi)^r \mu \approx \phi \;, \label{eq:Grover-lite}
\eeq for some integer $r$ to be determined, where ``$\approx$"
means approximation at large $\ns$. Thus, starting with an $\nb$
qubit system in a state $\mu$, one applies the operator $(-R_\mu
R_\phi)$ consecutively $r$ times, so that the $\nb$ qubit system
ends in a state as close to $\phi$ as possible. Measuring state
$\phi$ in the special basis yields the target state $j_{targ}$.

Eq.(\ref{eq:Grover-lite}) can be represented by a qubit circuit
consisting of a single wire for $\vec{\kappa}$, with $r$ nodes,
each representing $-R_\mu R_\phi$. Eq.(\ref{eq:Grover-lite}) can
also be represented by a QB net defined by a Markov chain graph
$\rvX_0\rarrow\rvX_1\rarrow \rvX_2 \rarrow \ldots\rarrow
\rvX_{r-1}$, with

\BeginQBNetTabular      \label{tab:Grover-qbnet}
    $\rvX_0$ & $X_0\in Bool^\nb$ & $\delta(X_0,0)$ & \\
\hline
    $\rvX_i$ for $i\in Z_{1,r-1}$ & $X_i \in Bool^\nb$ &
    $\bra{X_i} (-R_\mu R_\phi)\ket{X_{i-1}} $ & \\
\EndQBNetTabular

To find the optimum number $r$ of iterations,
one can proceed as follows.

First, notice that Eq.(\ref{eq:Grover-lite}) describes a process
which is entirely confined to the vector subspace spanned by $\mu$
and $\phi$. Since $\mu$ and $\phi$ are not orthogonal, it is
convenient to define an orthonormal basis $e_0, e_1$ for the space
$span(\mu, \phi)$. Let

\beq
e_0 = \phi
\;,
\;\;
e_1 = \frac{\phi^{not}}{\sqrt{\ns-1}}
\;.
\eeq
Then

\beq
\mu=\frac{1}{\sqrt{\ns}} (e_0 + \sqrt{\ns - 1}\;
e_1)
\;.
\eeq

\begin{figure}[h]
    \begin{center}
    \epsfig{file=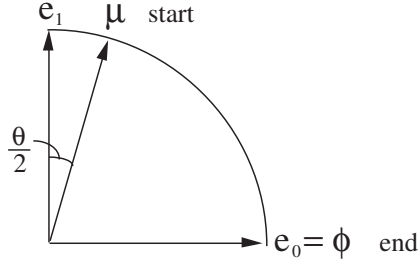, height=1.5in}
    \caption{Various vectors relevant to
Grover's Algorithm.}
    \label{fig:Grover-rot}
    \end{center}
\end{figure}

Fig.\ref{fig:Grover-rot} portrays various vectors that arise in
explaining Grover's algorithm.

Since we plan to stay within
the two dimensional vector space
with orthonormal basis $e_0, e_1$,
it is convenient to switch matrix representations.
Within $span(e_0, e_1)$,
$e_0, e_1$ can be represented more simply by:

\beq
e_0 = \left( \begin{array}{c} 1\\0 \end{array}
\right)
\;,\;\;
e_1 = \left( \begin{array}{c} 0\\1 \end{array}
\right)
\;.
\eeq
If
 $e_0, e_1$ are represented in this way, then

\beq \phi = \left( \begin{array}{c} 1\\0 \end{array} \right)
\;,\;\; \mu = \frac{1}{\sqrt{\ns}} \left( \begin{array}{c}
1\\\sqrt{\ns-1} \end{array} \right) \;,
\label{eq:Grover-2d-phi-mu} \eeq and

\beq
-R_\mu R_\phi =
\frac{1}{\ns}
\left(
\begin{array}{cc}
\ns-2 & 2 \sqrt{\ns-1} \\
-2 \sqrt{\ns-1} & \ns-2
\end{array}
\right)
\;.
\eeq
Thus,

\beq
-R_\mu R_\phi =
\left(
\begin{array}{cc}
\cos\theta & \sin\theta \\
-\sin\theta & \cos\theta
\end{array}
\right) \;, \label{eq:double-reflection-Grover} \eeq where

\beq \sin \theta = \frac{2\sqrt{\ns-1}}{\ns} \approx
\frac{2}{\sqrt{\ns}} \;. \eeq
Eq.(\ref{eq:double-reflection-Grover}) is just
 Eq.(\ref{eq:double-reflection-gen}) with
$e'_1=\mu$ and $e_0=\phi$.
It follows that

\beq
(-R_\mu R_\phi)^r =
\left(
\begin{array}{cc}
\cos(r\theta) & \sin(r\theta) \\
-\sin(r\theta) & \cos(r\theta)
\end{array}
\right)
\;,
\eeq
and

\beqa
(-R_\mu R_\phi)^r \mu &=&
\left(
\begin{array}{cc}
\cos(r\theta) & \sin(r\theta) \\
-\sin(r\theta) & \cos(r\theta)
\end{array}
\right)
\frac{1}{\sqrt{\ns}}
\left( \begin{array}{c} 1\\\sqrt{\ns-1}
\end{array} \right)\nonumber\\
&\approx &
\left( \begin{array}{c} \sin(r\theta) \\
\cos(r\theta) \end{array} \right)
\;.
\eeqa
We want the final state of the system to
be parallel or anti-parallel to $e_0=\phi$ ;
therefore, we want

\beq
\left( \begin{array}{c} \sin(r\theta) \\
\cos(r\theta) \end{array} \right)
\approx
\left( \begin{array}{c} \pm 1 \\ 0 \end{array} \right)
\;.
\eeq
This will occur if

\beq
r\theta \approx \frac{\pi}{2}(1+2k)
\;\;,\;\;
r\approx \frac{\pi}{4}(1+2k)\sqrt{\ns}
\eeq
for some integer $k$.

Note that, in Grover's algorithm,
the number of ``queries"
(calls to a unitary matrix
that depends on $\phi$) is
far from unique.
To illustrate this,
let $Q$ be a permutation matrix that
satisfies

\beq
Q \phi = \ket{0} = [1,0,0, \ldots, 0]^T
\;.
\eeq
Since all the components of $\mu$ are
equal, $Q\mu=\mu$. Thus

\beq
(-R_\mu R_\phi)^r \mu =Q^T (-R_\mu  R_{\ket{0}})^r Q
\mu =
Q^T (-R_\mu  R_{\ket{0}})^r  \mu
\;.
\eeq
Hence, it is possible to accomplish
the full Grover transformation of $\mu$
with only a single query $Q^T$.

Since
$\left(
\begin{array}{cc}
0 & 1\\
-1 & 0
\end{array}
\right)
\left(\begin{array}{c}a\\b\end{array}\right)
=
\left(\begin{array}{c}b\\-a\end{array}\right)$,
the matrix $\left(\begin{array}{cc}
0 & 1\\
-1 & 0
\end{array}\right)$ is just a clockwise
rotation by
$\pi/2$.
Let

\beqa
U_{Grov} &=&
\left(\begin{array}{cc}
0 & 1\\
-1 & 0
\end{array}\right)\nonumber\\
&=&-e_1 e_0^T + e_0 e_1^T\nonumber \\
&=& \frac{1}{\sqrt{\ns-1}}
[-\phi^{not} \phi^T + \phi (\phi^{not})^T]
\;.
\eeqa
Note that

\beqa
U_{Grov}\mu
&=& \frac{1}{\sqrt{\ns-1}}
\left(-\phi^{not} [\phi^T\mu] +
\phi [(\phi^{not})^T\mu]\right)
\nonumber\\
&=&
\frac{1}{\sqrt{\ns(\ns-1)}}
[-\phi^{not} + (\ns-1) \phi]
\;.
\eeqa
From the point of view of quantum compiling,
what Grover found is that
the $\pi/2$
rotation $U_{Grov}$
is (approximately) equal to
the $r$-fold product of $-R_\mu R_\phi$,
where $-R_\mu R_\phi$ can be shown to have a
SEO of low
(polynomial in $\nb$) complexity.

Grover's algorithm has been modified
in various, minor ways since it
was first published. For example,
Brassard et al. pointed out
in Ref.\cite{amplit-amplif}  that
the vector $\mu$ need not
be the vector whose  components are all equal.
Other vectors $\mu$ will do just as well.
Another modification of Grover's
algorithm due to Younes-Miller\cite{Younes}
adds an extra
qubit to the original $\nb$ qubits.
Next we will discuss
the Younes-Miller modification of Grover's algorithm,
because it
resembles a modification of Grover's algorithm that
 we will use in a future section.

Let $\vec{\kappa}=(\kappa_0, \kappa_1, \ldots,
 \kappa_{\nb-1})$
label $\nb$ bits.
Let $\tau$ label a single bit.
Assume $\tau$ and all the $\kappa_i$ are distinct.
Let $\mu$ and $\phi$ denote the same $\ns$
dimensional column vectors that we used
in discussing the original Grover algorithm.
In addition, define
 the following $2\ns$ dimensional column vectors:

\beq
\ket{\tmu} =
\ket{+_z}_\tau \ket{\mu}_{\vec{\kappa}} =
\left(
\begin{array}{c}1\\0\end{array}
\right)
\otimes \mu_\ns =
\left(
\begin{array}{c}\mu_\ns\\0\end{array}
\right)
\;,
\eeq

\beq
\ket{\tphi} =
\ket{-_x}_\tau \ket{\phi}_{\vec{\kappa}} =
\frac{1}{\sqrt{2}}
\left(
\begin{array}{c}1\\-1\end{array}
\right)
\otimes \phi =
\frac{1}{\sqrt{2}}
\left(
\begin{array}{c}\phi\\-\phi\end{array}
\right)
\;.
\eeq
Note that $\av{\tphi | \tmu} =
\frac{1}{\sqrt{2\ns}}$, so
$\tphi$ and $\tmu$ are nearly orthogonal
for large $\ns$.
Define projection and reflection operators for
$\tphi$ in the usual way:

\beq
\Pi_{\tphi} = \ket{\tphi}\bra{\tphi}
\;,\;
R_{\tphi} = 1-2\Pi_{\tphi}=
1 - 2
\Pi_{\ket{\phi}_{\vec{\kappa}}}\Pi_{\ket{-_x}_\tau
}
\;.
\eeq
$R_{\tphi}$ can be re-expressed
as

\beqa
R_{\tphi} &= &1 +
\Pi_{\ket{\phi}_{\vec{\kappa}}}(\sigma_x(\tau)
-1)=
\exp[{\Pi_{\ket{\phi}_{\vec{\kappa}}}\ln
\sigma_x(\tau)}]=\nonumber\\
&=&[\sigma_x(\tau)]^{\Pi_{\ket{\phi}_{\vec{\kappa}
}}}
\;.
\eeqa
Define projection and reflection operators for
$\tmu$ in the usual way:

\beq
\Pi_{\tmu} = \ket{\tmu}\bra{\tmu}
\;,\;
R_{\tmu} = 1-2\Pi_{\tmu}=
1 - 2
\Pi_{\ket{\mu}_{\vec{\kappa}}}\Pi_{\ket{+_z}_\tau}
\;.
\eeq
$R_{\tmu}$ can be re-expressed
as

\beq
R_{\tmu}
=
\hat{H}_1(\vec{\kappa}) \left( 1- 2
\Pi_{\ket{0}_{\vec{\kappa}}}
\Pi_{\ket{0}_\tau} \right)
\hat{H}_1(\vec{\kappa})
\;.
\eeq
In analogy with the original Grover's algorithm,
the Younes-Miller version can be summarized
by

\beq
(-R_{\tmu}R_{\tphi})^r
\ket{\tmu} \approx \ket{\tphi}
\;,
\label{eq:grov-Younes}
\eeq
for some integer $r$ to be determined,
where ``$\approx$" means approximation at large $\ns$.
Thus, starting with
an $\nb+1$ qubit system in a state $\tmu$, one
applies the operator
$(-R_\mu R_\phi)$ consecutively $r$ times,
so that the final state
of the $\nb+1$ qubit system ends in a state
as close to $\tphi$
as possible. Measuring state $\tphi$ in
the special basis yields the target state $j_{targ}$.

To find the optimum number $r$ of iterations,
one can proceed as follows.

First, notice that Eq.(\ref{eq:grov-Younes})
describes a process which is entirely
confined to the vector
subspace spanned by $\tmu$ and $\tphi$.
Since $\tmu$ and $\tphi$ are not orthogonal,
it is convenient to define an orthonormal
basis $e_0, e_1$ for the space
$span(\tmu, \tphi)$. Let

\beq
e_0 = \tphi=
\frac{1}{\sqrt{2}}
\left(
\begin{array}{c}\phi\\-\phi\end{array}
\right)
\;,
\eeq
and

\beq
e_1 = \frac{1}{K}[\tmu - (\tmu\cdot
 e_0) e_0]
\;,
\eeq
where $K$ is chosen so that $e_1^2=1$.
It is easy to show that

\beq
K = |\tmu - (\tmu\cdot  e_0) e_0|=
\sqrt{ \frac{\ns - \frac{1}{2}}{\ns} }
\;.
\eeq
Thus,

\beq
e_1 = \frac{1}{\sqrt{\ns - \frac{1}{2}}}
\left(
\begin{array}{c}\phi^{not} + \frac{\phi}{2}\\
\frac{\phi}{2}\end{array}
\right)
\;.
\eeq
Furthermore,

\beq
\tmu =
\frac{1}{\sqrt{2\ns}}[ e_0 +
\sqrt{2\ns - 1}\; e_1 ]
\;.
\eeq

\begin{figure}[h]
    \begin{center}
    \epsfig{file=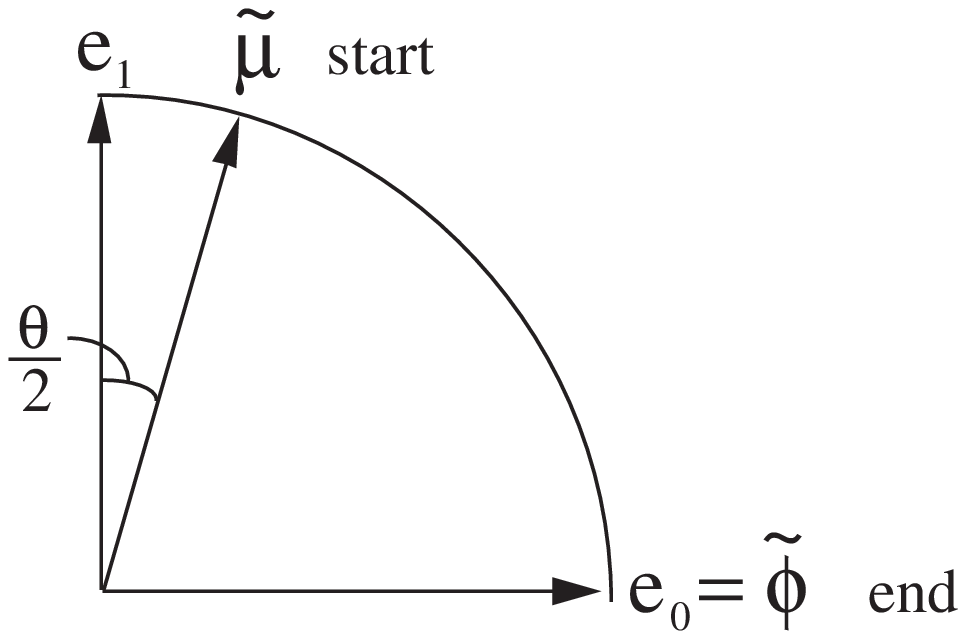, height=1.5in}
    \caption{Various vectors relevant to
    Younes' version of
Grover's Algorithm.}
    \label{fig:Younes-rot}
    \end{center}
\end{figure}

Fig.\ref{fig:Younes-rot} portrays various
vectors that arise in explaining
Younes' version of
Grover's algorithm.

Since we plan to stay within
the two dimensional vector space
with orthonormal basis $e_0, e_1$,
it is convenient to switch matrix representations.
Within $span(e_0, e_1)$,
$e_0, e_1$ can be represented more simply by:

\beq
e_0 = \left( \begin{array}{c} 1\\0 \end{array}
\right)
\;,\;\;
e_1 = \left( \begin{array}{c} 0\\1 \end{array}
\right)
\;.
\eeq
If $e_0, e_1$ are represented this way, then

\beq
\tphi= \left( \begin{array}{c} 1\\0
\end{array} \right)
\;,\;\;
\tmu = \frac{1}{\sqrt{2\ns}}
\left( \begin{array}{c} 1 \\
\sqrt{2\ns - 1} \end{array} \right) \;,
\label{eq:Younes-2d-phi-mu} \eeq and

\beq
-R_{\tmu} R_{\tilde{\phi }}=
\frac{1}{\ns}
\left(
\begin{array}{cc}
\ns-1 &  \sqrt{2\ns-1} \\
- \sqrt{2\ns-1} & \ns-1
\end{array}
\right)
\;.
\eeq
Thus,

\beq
-R_{\tmu} R_{\tilde{\phi }} =
\left(
\begin{array}{cc}
\cos\theta & \sin\theta \\
-\sin\theta & \cos\theta
\end{array}
\right)
\;,
\eeq
where

\beq
\sin \theta = \frac{\sqrt{2\ns-1}}{\ns}
\approx \sqrt{\frac{2}{\ns}}
\;.
\eeq

A comparison of Eq.(\ref{eq:Grover-2d-phi-mu}) (for the original
Grover's algorithm) and Eq.(\ref{eq:Younes-2d-phi-mu}) (for
Younes's version of Grover's algorithm) reveals that for the
purpose of finding the optimal number $r$ of iterations, Younes'
algorithm is the same as Grover's algorithm if one replaces
 $\ns$ in Grover's algorithm by $2\ns$.
 This comes from the fact that Younes'
 algorithm
uses $\nb + 1$ bits whereas Grover's uses
$\nb$.

\section{Generalization of Standard Algorithms,\\
 a list of Desiderata}

So far we have analyzed several
standard quantum computing
algorithms, namely those attributed to
Deutsch-Jozsa, Bernstein-Vazirani, Simon
and Grover.
(Two other standard algorithm's that
we didn't analyze are Shor's
 algorithm\cite{Shor} and the
algorithm for Teleportation\cite{Tele}.)
In this section, we will try to
point out those
features of the standard algorithms that
would be, in our opinion, fruitful to generalize.
Bear in mind that generalizations are seldom
unique,
but some are more natural, fruitful and
far-reaching than others.

\MyCases{(a) Allow more complicated graph
topologies}

The standard algorithms discussed here can all be
represented by
QB nets with trivial  topologies
such as 2 body scattering graphs or Markov chains.
However, other important quantum algorithms, such
as the one for Teleportation\cite{Tele},
can be represented by QB nets with  more
complicated
graph topologies (e.g., with  loops).

\MyCases{(b) Estimate more general probability
distributions}

The goal of most
 standard algorithms  is to estimate
a deterministic probability distribution.
However, estimating non-deterministic ones is
also very useful. Such estimates
are useful in, for example, applications of
Decision Theory and Artificial Intelligence,
where inferences are made based on uncertain
knowledge.

\MyCases{(c) Allow multiple runs and the rejection
of some}

If one is estimating a non-deterministic
probability distribution,
it will be necessary to do
multiple runs.
It may also be necessary to allow rejection of
runs.
Obviously, the number of rejected runs
is best kept as small as possible.

\MyCases{(d) Allow more general measurements}

Suppose $\rvx$ is a node of a QB net. Let $S_\rvx$
be the set of its possible states.
We will say that node  $\rvx$ has been
measured  if
during the experiment which the QB net describes,
a measurement is
performed  that restricts the
possible states of $\rvx$ to a proper
subset $S'_\rvx$ of $S_\rvx$.
When $\rvx$ is an internal
(ditto, external) node of the QB net, we
will refer to its measurement
as an internal (ditto, external) measurement.

The standard algorithms discussed here
use external but no internal
measurements.
However, other important quantum algorithms, such
as the one
for Teleportation, do use internal ones.

\section{Q-Embeddings}

The remainder of this paper will be
devoted to discussing a class
of algorithms which generalizes
some standard algorithms and
achieves some of the desiderata
given in the previous section.
Our algorithms are based on the idea that, given a
CB net, one can always embed it in a QB net.
Simple examples of such q-embeddings
 have already
been given in the
sections dealing with standard algorithms.

We start by defining some terminology that will
be useful.

A {\bf probability matrix} $P(y|x)$ is
a rectangular (not necessarily square)
matrix with row index $y\in S_\rvy$ and column
index $x\in S_x$ such that $P(y|x)\geq 0$ for all $x,y$,
and $\sum_y P(y|x) = 1$ for all $x$.
The set of all probability matrices $P(y|x)$
where $x\in S_\rvx$ and $y\in S_\rvy$
will be denoted by $pd(S_\rvy|S_\rvx)$
 (pd = probability
distribution).
A probability matrix is assigned to each node of a
CB net.
A {\bf probability matrix} $P(y|x)$ {\bf is
deterministic} if
for each column $x$, there exists a single row
$y$, call it $y(x)$, such that $P(y|x) =
\delta(y(x), y)$.
Any map $f:S_\rvx\rarrow S_\rvy$ uniquely
specifies (and is uniquely specified) by
the deterministic probability matrix $P$ with
matrix elements
$P(y|x) =  \delta(y, f(x))$ for all $x\in S_\rvx$
and $y\in S_\rvy$.
We will often talk about a map $f$ and its
associated probability matrix $P(y|x)$
as if they were the same thing.

Given two matrices $A$ and $B$ of the same dimensions,
their {\bf Hadamard product} $C= A\odot B$
is defined by $C_{i,j} = A_{i,j}B_{i,j}$
for all $i,j$. We will call
${\rm HAS}(A) = A\odot A^\dagger$ the
{\bf Hadamard Absolute Square (HAS) of matrix} $A$.
If $U$ is a unitary matrix, then ${\rm HAS}(U)$ is a
probability matrix.
For example, for any angle $\theta$,

\beq
{\rm HAS}(
\left [
\begin{array}{cc}
\cos\theta & \sin\theta \\
-\sin\theta& \cos\theta
\end{array}
\right]
) =
\left [
\begin{array}{cc}
\cos^2\theta & \sin^2\theta \\
\sin^2\theta & \cos^2\theta
\end{array}
\right]
\;.
\eeq
Another
 example is

\beq
{\rm HAS}(\hat{H}_1) =
\frac{1}{2}
\left [
\begin{array}{cc}
1 & 1 \\
1 & 1
\end{array}
\right]
\;.
\eeq

A CB net $\cbnet$ is the {\bf HAS of QB net}
$\qbnet$
if $\qbnet$ and $\cbnet$ have the same graph, and
their
node matrices are related as follows.
For each node $\rvx_i$, if
$A[x_i | (x.)_{\Gamma_i}]$ is the amplitude
of node $\rvx_i$ in $\qbnet$,
and $P[x_i | (x.)_{\Gamma_i}]$ is the probability
of node $\rvx_i$ in $\cbnet$, then
$|A[x_i | (x.)_{\Gamma_i}]|^2 = P[x_i | (x.)_{\Gamma_i}]$.
In such a case, we will write ${\rm
HAS}(\qbnet) = \cbnet$.

A unitary matrix $A(y, \tilde{x} | x, \tilde{y})$
(with rows labelled
by $y, \tilde{x}$ and columns by $x, \tilde{y}$)
is a {\bf q-embedding of probability
matrix} $P(y|x)$ if

\beq
\sum_{\tilde{x}} | A(y, \tilde{x} | x,
\tilde{y}=0) |^2 = P(y|x)
\;
\label{eq:q-embed-mat-def}\eeq
for all possible values of $y$ and $x$. (the ``q" in
``q-embedding" stands for ``quantum"). We say
$\tilde{y}$ is a {\bf source index} and
$\tilde{x}$ is a {\bf sink index}.
We also refer to $\tilde{x}$ and $\tilde{y}$
collectively as
{\bf ancilla indices}.
Note that any unitary matrix is a q-embedding
 of its HAS.
Indeed,
in this case Eq.(\ref{eq:q-embed-mat-def})
is satisfied with the indices $\tilde{x}$ and
$\tilde{y}$
each ranging over a single value
(i.e., $\tilde{x}$ and $\tilde{y}$ are fixed).
If a q-embedding satisfies
$A(y, \tilde{x} | x, \tilde{y})\in Bool$
for all $y, \tilde{x}, x, \tilde{y}$,
we say that it is a
{\bf deterministic q-embedding or a
deterministic reversible extension (DRE)
of its probability matrix} (note that its
 probability matrix must also be
deterministic).
By an extension of a matrix we
mean adding extra rows and/or columns to it.
General q-embeddings use the square
root of the entries of the
original probability matrix
so they are not simply extensions
of the original matrix; they are, however,
reversible since they are unitary matrices.

Given a QB net $\qbnet$,
let

\beq
P[ (x.)_L] =
\left| \sum_{(x.)_{\Gamma_Q-L}}
A(x.)
\right|^2
\;.
\label{eq:q-embed-net-predef}\eeq
On the right hand side of
Eq.(\ref{eq:q-embed-net-predef}),
$A(x.)$ is the amplitude of story $(x.)$,
$\Gamma_Q$  is the set of indices
of all the
nodes of $\qbnet$,
and
$L$ is the set of indices of
 all leaf (aka external) nodes
of $\qbnet$.
We say $\qbnet$ is a {\bf q-embedding
of CB net} $\cbnet$ if
$P[ (x.)_L]$ defined by
Eq.(\ref{eq:q-embed-net-predef}) satisfies

\beq
P[ (x.)_{\Gamma_C}]
=
\sum_{L_1} P[ (x.)_L]
\;,
\label{eq:q-embed-net-def}
\eeq
where $L_1\subset L$,
and $\Gamma_C$ is the set of
indices of all nodes of $\cbnet$.
Thus, the probability distribution associated
with all nodes of $\cbnet$ can be obtained
from the probability distribution associated
with the external nodes of $\qbnet$.
Some examples of q-embeddings of CB nets
have already been given during our
discussion of standard algorithms.
More examples will be given in subsequent sections.

For some positive integers $r$ and $s$,
we will say a map $f:Bool^r\rarrow Bool^s$
is a {\bf binary
gate} from $r$ to $s$ bits.
$f$ uniquely specifies (and is uniquely specified)
by the deterministic probability matrix
with entries $P(y|x) =
\delta(f(x), y)$,
where
$x=(x_0, x_1, \dots, x_{r-1})\in Bool^r$ and
$y = (y_0, y_1, \dots, y_{s-1})\in Bool^s$.
If $f$ is an invertible map, we will say that the
gate is {\bf reversible}.
For example, the AND gate
which takes $(x_1, x_0)\rarrow y_0$
with $y_0 = x_0 x_1$ is a binary gate.
So are the OR and NOT gates. Out of these 3 gates,
only the NOT gate is reversible.

\begin{figure}[h]
    \begin{center}
    \epsfig{file=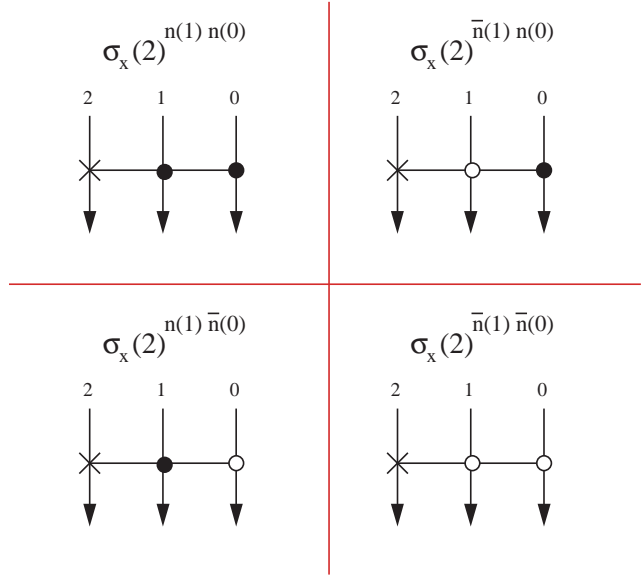, height=3in}
    \caption{Four different kinds of Toffoli
gates. 0,1,2 are bit labels.}
    \label{fig:Toffoli1}
    \end{center}
\end{figure}

Another example of a reversible binary gate is the
{\bf Toffoli gate}\cite{F-T}.
It maps 3 bits into 3 bits as follows:

\beq
\begin{array}{l}
y_0 = T_0(x) = x_0, \\
y_1 = T_1(x) =x_1, \\
y_2 = T_2(x) = x_2 \oplus x_0 x_1.
\end{array}
\;
\eeq
The Toffoli gate can  also be defined as the
following  deterministic probability matrix

\beq
P(y|x) = \delta(y, T(x)) = \delta(y_2, x_2 \oplus
x_0 x_1) \delta(y_1, x_1) \delta(y_0, x_0)
\;.
\label{eq:toffoli-def-del}\eeq
Consider
3 bits labelled 0, 1, and 2, and suppose
the $i$th bit changes value from
$x_i$ to $y_i$.
Then bits 0 and 1 do not change whereas bit
2 flips iff the product $x_0x_1$ equals one.
Thus,  the probability
matrix with entries given by
Eq.(\ref{eq:toffoli-def-del}) is simply
a doubly controlled not:

\beq
[P(y|x)] = \sigma_x(2)^{\nop(1)\nop(0)}
\;.
\label{eq:toffoli-def-sig}\eeq
It is convenient to use the term
Toffoli gate to refer not only to the gate defined
by Eq.(\ref{eq:toffoli-def-del}), but also to
the 3 other gates that
one obtains by replacing $x_0x_1$ in
Eq.(\ref{eq:toffoli-def-del}) by
$x_0\overline{x_1}$,
or $\overline{x_0}x_1$,
or $\overline{x_0}\overline{x_1}$.
This corresponds to replacing
$\nop(1)\nop(0)$ in Eq.(\ref{eq:toffoli-def-sig})
by
$\nop(1)\nbar(0)$,
or $\nbar(1)\nop(0)$,
or $\nbar(1)\nbar(0)$.
Fig.\ref{fig:Toffoli1} shows the 4
doubly-controlled nots
that we call Toffoli gates as well as the circuit
diagrams
usually used to represent them.

\begin{figure}[h]
    \begin{center}
    \epsfig{file=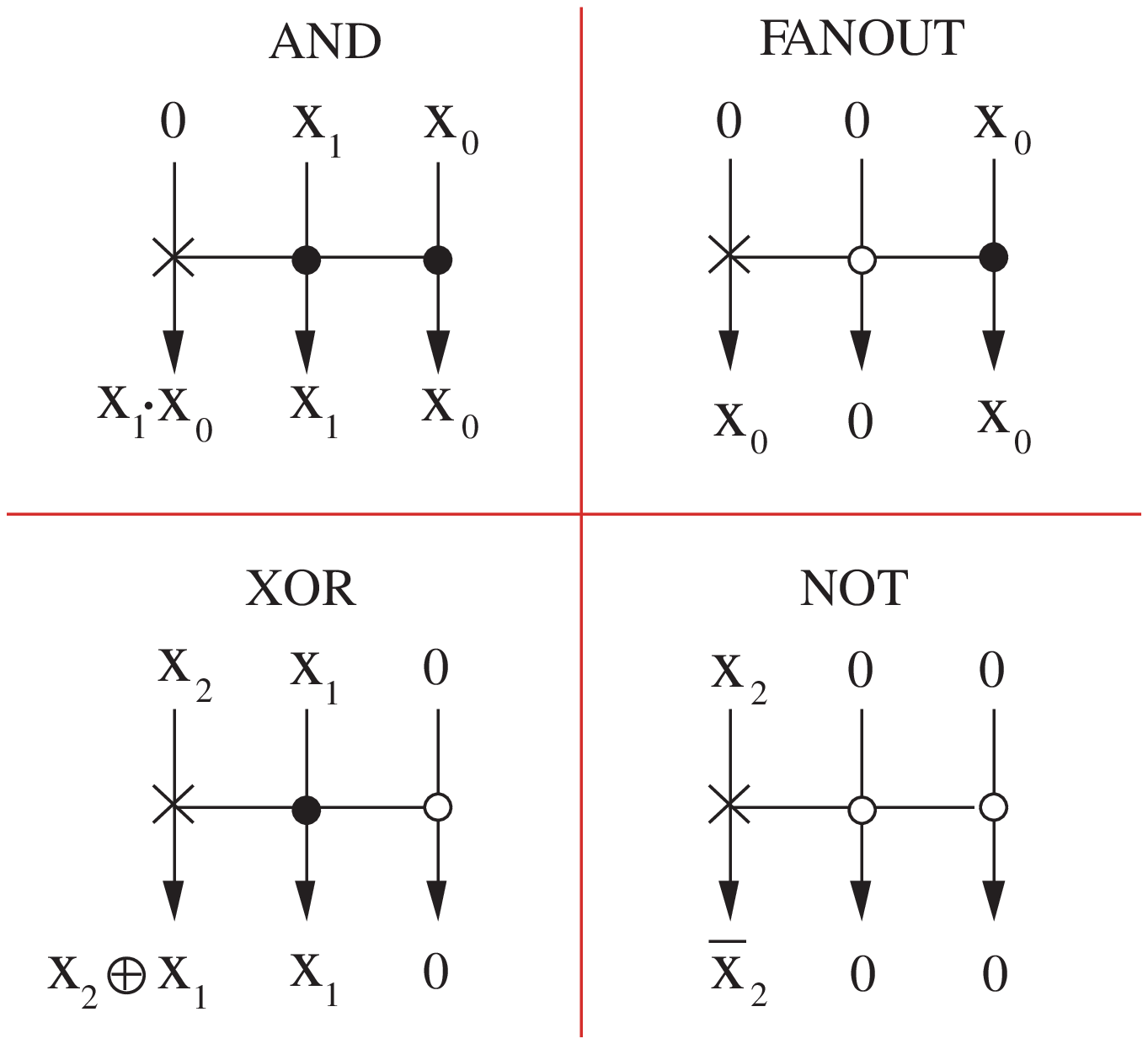, height=3in}
    \caption{How to express elementary gates in
terms of  Toffoli  gates. $x_0, x_1, x_2\in
Bool$ are bit values.}
    \label{fig:Toffoli2}
    \end{center}
\end{figure}

\subsection{Q-Embedding of Probability Matrices}

In this section we will first give some
examples of q-embeddings of
probability matrices.
Then we will show that any probability matrix has
a q-embedding.

Any unitary matrix is a q-embedding of
its HAS, but such q-embeddings
are trivial in the sense that they have no ancilla
indices.

As first shown in Refs.\cite{F-T},
 the Toffoli gates
can be used to build
q-embeddings (in fact,
 DREs) of the
elementary binary gates AND, XOR, NOT, FANOUT. See
Fig.\ref{fig:Toffoli2}.
Let $x=(x_0, x_1, x_2)\in Bool^3$
and $y=(y_0, y_1, y_2)\in Bool^3$.
For the AND gate,

\begin{subequations}
\label{eq:q-embed-of-elem-gates}
\beq
\sum_{y_1, y_0} \left| \av{y |
\sx(2)^{\nop(1)\nop(0)} | x_2 = 0, x_1, x_0}
\right |^2 =
\delta(y_2, x_1 x_0)
\;.
\eeq
For the FANOUT gate,

\beq
\sum_{y_1} \left| \av{y | \sx(2)^{\nbar(1)\nop(0)}
| x_2=0, x_1=0, x_0} \right |^2 =
\delta(y_2, x_0) \delta(y_0, x_0)
\;.
\eeq
For the XOR gate,

\beq
\sum_{y_1, y_0} \left| \av{y |
\sx(2)^{\nop(1)\nbar(0)} | x_2, x_1, x_0=0} \right
|^2 =
\delta(y_2, x_2 \oplus x_1)
\;.
\eeq
For the NOT gate,

\beq
\sum_{y_1, y_0} \left| \av{y |
\sx(2)^{\nbar(1)\nbar(0)} | x_2, x_1=0, x_0=0}
\right |^2 =
\delta(y_2, x_2 \oplus 1) = \delta(y_2,
\overline{x_2})
\;.
\label{eq:not-gate}\eeq
\end{subequations}
Note that the NOT gate is just $\sx$, which is a
DRE of itself.
Eq.(\ref{eq:not-gate}) gives a different
DRE of $\sx$.
In the left hand side of
Eqs.(\ref{eq:q-embed-of-elem-gates}), the $x_i$
indices
that are set to zero are called {\bf source
indices}, and the $y_i$ indices
that are summed over are called {\bf sink
indices}.
Sink and source indices are collectively called
{\bf ancilla indices}.

Next we will prove that any probability matrix has
a q-embedding.
Suppose that we are given a probability matrix
$P(y|x)$
where $x\in S_\rvx$ and $y\in S_\rvy$. Let
$N_\rvx$ (ditto, $N_\rvy$) denote
the number of elements in $S_\rvx$ (ditto,
$S_\rvy$).
Let $\xi^{(x)}$ for $x\in S_\rvx$ be any
orthonormal basis of the complex $N_\rvx$
dimensional vector space.
The components of $\xi^{(x)}$ will
be denoted by $\xi^{(x)}_{\tilde{x}}$, where
$\tilde{x}\in S_\rvx$.
If the $\xi^{(x)}$'s are the standard
basis,
then $\xi^{(x)}_{\tilde{x}} = \delta(x,
\tilde{x})$.
Define matrix $A$ by

\beq
A(y, \tilde{x} | x, \tilde{y}) =
\left \{
\begin{array}{ll}
\sqrt{P(y|x)} \;\;\xi^{x}_{\tilde{x}} & \mbox{  if}\;
\tilde{y} = 0 \\
\mbox{obtained by Gram-Schmidt method} & \mbox{  if}\;
\tilde{y} \neq 0
\end{array}
\right .
\;.
\label{eq:gen-mat-q-embed}\eeq

\begin{figure}[h]
    \begin{center}
    \epsfig{file=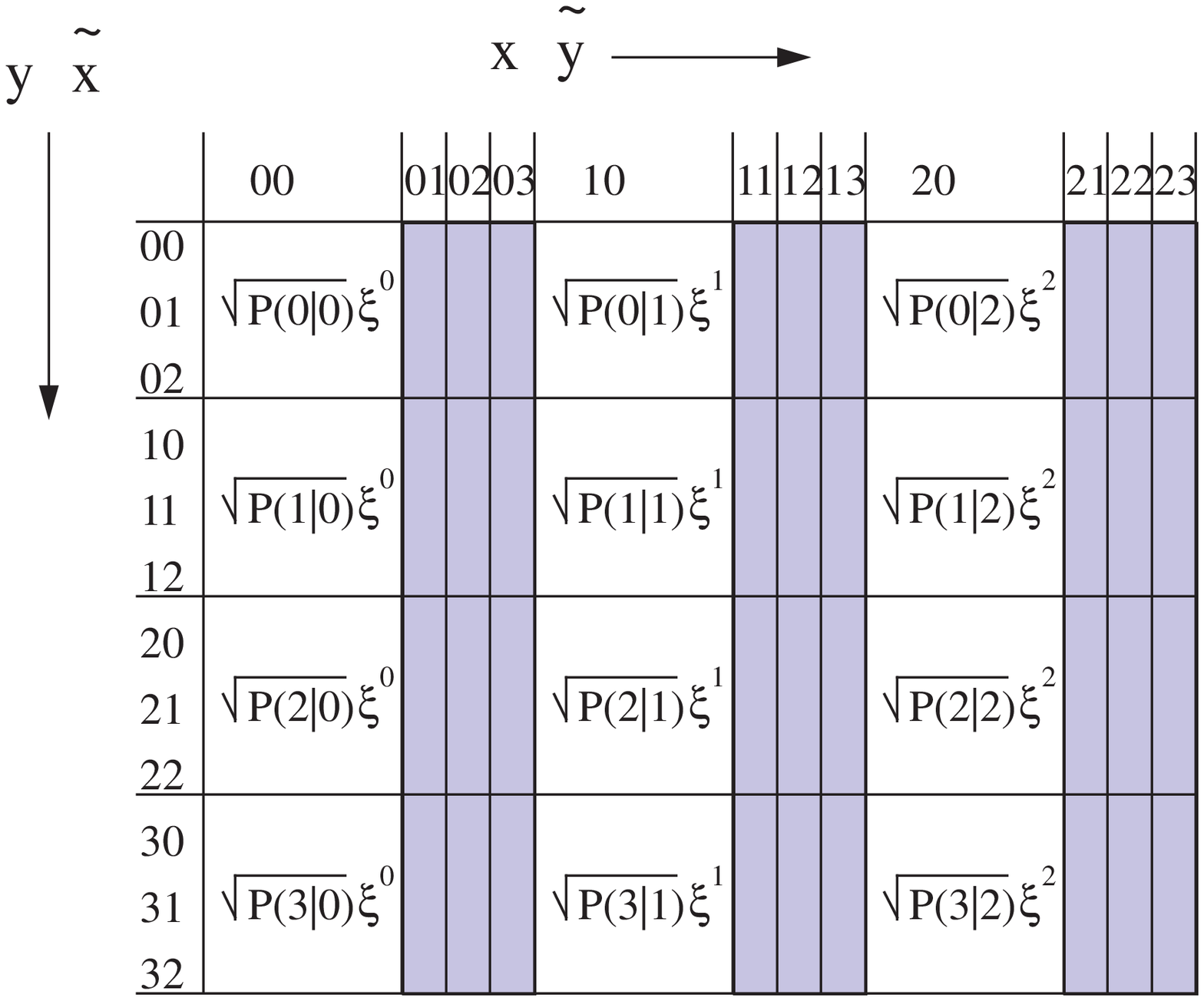, width=5in}
    \caption{How to construct a q-embedding
     of any probability matrix.}
    \label{fig:mat-q-embed}
    \end{center}
\end{figure}

To understand the last equation, consider
Fig.\ref{fig:mat-q-embed}.
 In that figure we have assumed
for definiteness that $S_\rvx = \{0,1,2\}$ and
$S_\rvy = \{0,1,2,3\}$. The shaded (ditto,
unshaded) columns  have $\tilde{y} \neq 0$ (ditto,
$\tilde{y} = 0$).
It is easy to see that the unshaded columns are
orthonormal
because the vectors $\xi^{x}$ are orthonormal and
$\sum_y P(y|x) = 1$.
Since the unshaded columns are orthonormal, one
can use  the Gram-Schmidt
method\cite{Noble} to fill the shaded columns so
that all the columns of $A$
are orthonormal and therefore $A$ is unitary. Note
that by virtue
of Eq.(\ref{eq:gen-mat-q-embed}),

\beq
\sum_{\tilde{x}}
|A(y, \tilde{x} | x, \tilde{y}=0 ) |^2 =
\sum_{\tilde{x}} P(y|x) \xi^{(x)*}_{\tilde{x}}
\xi^{(x)}_{\tilde{x}} = P(y|x)
\;
\eeq
so that the $A$ defined by
Eq.(\ref{eq:gen-mat-q-embed}) does indeed
satisfy
Eq.(\ref{eq:q-embed-mat-def}).\cite{real-extension}

Note that the matrix $A$ defined by
Eq.(\ref{eq:gen-mat-q-embed})
has dimensions $N_\rvx N_\rvy \times  N_\rvx
N_\rvy$.
It is sometimes possible to find a
smaller q-embedding of an $N_\rvy \times N_\rvx$
probability matrix $P(y|x)$.
For example, $\sigma_x$ is a
q-embedding of itself. As a less trivial
example, suppose

\beq
P(y|x_1, x_2) = \delta(y, x_1 \oplus x_2)
\;,
\eeq
for $y, x_1, x_2\in Bool$. Then define

\beq
A(y, e | x_1, x_2) = \frac{(-1)^{x_1
e}}{\sqrt{2}}\delta(y, x_1 \oplus x_2)
\;,
\eeq
for $y, e, x_1, x_2\in Bool$.
It is easy to check that matrix $A$ is unitary.
Furthermore,

\beq
\sum_e |A(y, e | x_1, x_2) |^2 =
\frac{1}{2} \sum_e \delta(y, x_1 \oplus x_2) =
\delta(y, x_1 \oplus x_2)
\;.
\eeq

\subsection{Q-Embedding of CB Nets}

As we've said before,
F-T showed in Refs.\cite{F-T}
how,  given any
binary gate $f$, one can construct another
binary gate $\overline{f}$
such that $\overline{f}$ is a
DRE of $f$.
Their method for constructing $\overline{f}$ is
to first represent $f$ as a binary deterministic
circuit composed of elementary gates
(AND, XOR, NOT, FANOUT), and then
to  modify the circuit by replacing each of its gates
by  a DRE of it.
The desired gate $\overline{f}$ is then specified by
the modified circuit.

 In this section we will show how, given
 any CB net $\cbnet$, one can construct a QB net
 $\qbnet$ which is a q-embedding of $\cbnet$.
So far we've shown how to construct a q-embedding for
any probability matrix. Now remember that each
node of $\cbnet$ has a probability
matrix assigned to it.
The main step in constructing
a q-embedding of $\cbnet$ is to
replace each node matrix of $\cbnet$ with
a q-embedding of it.
Thus, our method
for constructing a q-embedding
of a CB net is
a generalization of the  F-T method for constructing
a DRE of a binary deterministic circuit. We
generalize their method so that it
can be applied to any
classical stochastic circuit,
 not just
binary deterministic ones.

Before describing our construction method,
we need some definitions.
We say a node $\rvm$
 is a {\bf marginalizer node} if it
has a single input arrow and a single
output arrow.
Furthermore, the parent node of $\rvm$,
call it $\rvx$,
has states $x=(x_1, x_2, \ldots, x_n)$,
where  $x_i\in S_{\rvx_i}$ for each $i\in Z_{1,n}$.
Furthermore, for some particular integer
$i_0\in Z_{1,n}$,
the set of possible states of $\rvm$
is $S_\rvm= S_{\rvx_{i_0}}$, and
the node matrix of $\rvm$ is
$P(\rvm=m|\rvx=x)=\delta(m,x_{i_0})$.

Let $\cbnet$ be a CB net for which we want to
obtain a q-embedding.
Our construction has two steps:

\MyCases{(Step 1) Add marginalizer nodes.}

More specifically, replace $\cbnet$ by a
modified  CB net $\cbnet_{mod}$ obtained
 as follows.
For each node $\rvx$ of $\cbnet$,
add a marginalizer node between $\rvx$
and every child of $\rvx$. If $\rvx$ has no
children, add a child to it.

\begin{figure}[h]
    \begin{center}
    \epsfig{file=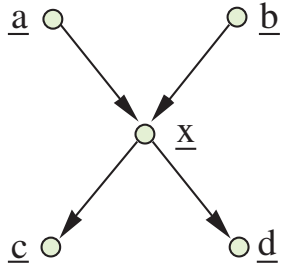, height=1.5in}
    \caption{
CB net for 2-body scattering.
We show how to construct a q-embedding for this
CB net.}
    \label{fig:scat-cbnet}
    \end{center}
\end{figure}

As an example of this step, consider the
net $\cbnet$ (``two body scattering net")
defined by Fig.\ref{fig:scat-cbnet}
and Table \ref{tab:scat-cbnet}.

\BeginCBNetTabular      \label{tab:scat-cbnet}
    $\rva$ & $a\in S_\rva$ & $P(a)$ &\\
\hline
    $\rvb$ & $b\in S_\rvb$ & $P(b)$ & \\
\hline
    $\rvc$ & $c\in S_\rvc$ & $P(c|x)$ & \\
\hline
    $\rvd$ & $d\in S_\rvd$ & $P(d|x)$ & \\
\hline
    $\rvx$ & $x\in S_\rvx$ & $P(x|a,b)$ & \\
\EndQBNetTabular

\begin{figure}[h]
    \begin{center}
    \epsfig{file=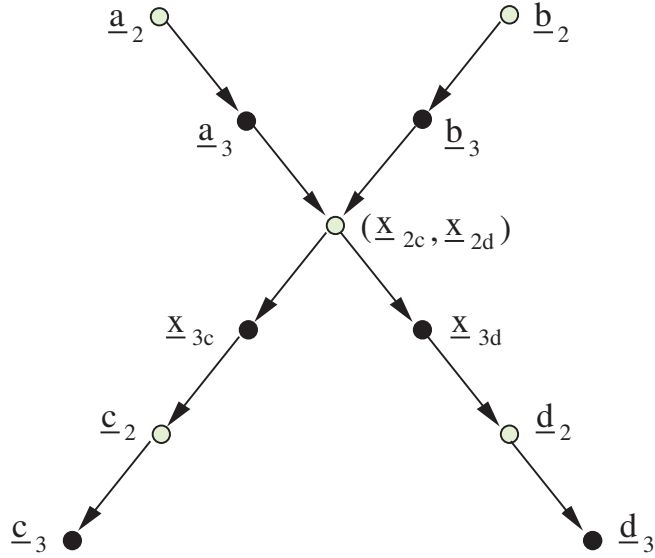, height=3in}
    \caption{CB net of Fig.\ref{fig:scat-cbnet}
     after adding marginalizer nodes.}
    \label{fig:scat-mod-cbnet}
    \end{center}
\end{figure}

Applying Step 1 to
$\cbnet$ for two body scattering yields
$\cbnet_{mod}$ defined by
 Fig.\ref{fig:scat-mod-cbnet} and
 Table \ref{tab:scat-mod-cbnet}.

\BeginCBNetTabular      \label{tab:scat-mod-cbnet}
    $\rva_2$ & $a_2\in S_\rva$ & $P_\rva(a_2)$ &\\
\hline
    $\rva_3$ & $a_3\in S_\rva$ & $\delta(a_3, a_2)$ &\\
\hline
    $\rvb_2$ & $b_2\in S_\rvb$ & $P_\rvb(b_2)$ &\\
\hline
    $\rvb_3$ & $b_3\in S_\rvb$ & $\delta(b_3, b_2)$ &\\
\hline
    $\rvc_2$ & $c_2\in S_\rvc$
    & $P_{\rvc|\rvx}(c_2|x_{3c})$ & \\
\hline
    $\rvc_3$ & $c_3\in S_\rvc$
    & $\delta(c_3, c_2)$ & \\
\hline
    $\rvd_2$ & $d_2\in S_\rvd$
    & $P_{\rvd|\rvx}(d_2|x_{3d})$ & \\
\hline
    $\rvd_3$ & $d_3\in S_\rvd$
    & $\delta(d_3, d_2)$ & \\
\hline
    $(\rvx_{2c}, \rvx_{2d})$ & $(x_{2c}, x_{2d})\in S_\rvx^2$ &
    $P_{\rvx|\rva,\rvb}(x_{2c}|a_3,b_3)\delta(x_{2d}, x_{2c})$ & \\
\hline
    $\rvx_{3c}$ & $x_{3c}\in S_\rvx$ &
    $\delta(x_{3c}, x_{2c})$ & \\
\hline
    $\rvx_{3d}$ & $x_{3d}\in S_\rvx$ &
    $\delta(x_{3d}, x_{2d})$ & \\
\EndQBNetTabular

\MyCases{(Step 2) Replace node probability matrices
by their q-embeddings. Add ancilla nodes.}

More
specifically, replace $\cbnet_{mod}$
by a QB net $\qbnet$ obtained as follows.
For each
node of $\cbnet_{mod}$, except for
the marginalizer nodes that were
added in the previous step,
replace its node matrix by a
new node matrix which is a
q-embedding of the original node matrix.
Add a new node
for each ancilla index of
each new node matrix.
These new nodes will be called
{\bf ancilla nodes} (of either
the source or sink type) because
they correspond to ancilla indices.

\begin{figure}[h]
    \begin{center}
    \epsfig{file=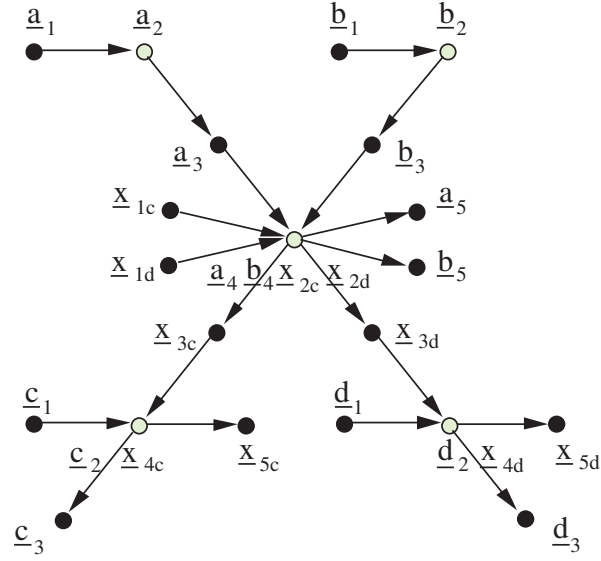, height=3in}
    \caption{A QB Net which is a q-embedding
    for the CB net of Fig.\ref{fig:scat-cbnet}.}
    \label{fig:scat-qbnet}
    \end{center}
\end{figure}

Applying Step 2 to net
$\cbnet_{mod}$ for two body scattering yields
$\qbnet$ defined by Fig.\ref{fig:scat-qbnet}
and Table \ref{tab:scat-qbnet}.

{\footnotesize
\BeginQBNetTabular      \label{tab:scat-qbnet}
    $\rva_1$ & $a_1\in S_\rva$ & $\delta(a_1, 0)$&\\
\hline
    $\rva_2$ & $a_2\in S_\rva$ &
    $A(a_2|a_1=0)=\sqrt{P_\rva(a_2)}$ &\\
\hline
    $\rva_3$ & $a_3\in S_\rva$ & $\delta(a_3, a_2)$\\
\hline
    $(\rva_4,\rvb_4,\rvx_{2c},\rvx_{2d}) $
    & $(a_4,b_4,x_{2c},x_{2d})
    \in S_{\rva, \rvb, \rvx,\rvx}$
    & $A(a_4, b_4, x_{2c}, x_{2d}|a_3, b_3, x_{1c}=0,
    x_{1d}=0)=$ & \\
    &&$\sqrt{P_{\rvx|\rva,\rvb}(x_{2c}|a_3, b_3)}
    \delta(a_4, a_3)\delta(b_4, b_3)\delta(x_{2d},
    x_{2c}) $&\\
\hline
    $\rva_5$ & $a_5\in S_{\rva}$
    & $\delta(a_5, a_4)$ & \\
\hline
    $\rvb_1$ & $b_1\in S_\rvb$ & $\delta(b_1, 0)$&\\
\hline
    $\rvb_2$ & $b_2\in S_\rvb$ &
    $A(b_2|b_1=0)=\sqrt{P_\rvb(b_2)}$ &\\
\hline
    $\rvb_3$ & $b_3\in S_\rvb$ & $\delta(b_3, b_2)$\\
\hline
    $\rvb_5$ & $b_5\in S_{\rvb}$
    & $\delta(b_5, b_4)$ & \\
\hline
    $\rvc_1$ & $c_1\in S_\rvc$ & $\delta(c_1, 0)$&\\
\hline
    $(\rvc_2, \rvx_{4c})$ & $(c_2, x_{4c})\in
    S_{\rvc,\rvx}$ & $A(c_2, x_{4c}|c_1=0,
    x_{3c})=$&\\
    &&$\sqrt{P_{\rvc|\rvx}(c_2|x_{3c})}\delta(x_{4c},
    x_{3c})$&\\
\hline
    $\rvc_3$ & $c_3\in S_\rvc$ & $\delta(c_3,c_2)$ &\\
\hline
    $\rvd_1$ & $d_1\in S_\rvd$ & $\delta(d_1, 0)$&\\
\hline
    $(\rvd_2, \rvx_{4d})$ & $(d_2, x_{4d})\in
    S_{\rvd,\rvx}$ & $A(d_2, x_{4d}|d_1=0,
    x_{3d})=$&\\
    &&$\sqrt{P_{\rvd|\rvx}(d_2|x_{3d})}\delta(x_{4d},
    x_{3d})$&\\
\hline
    $\rvd_3$ & $d_3\in S_\rvd$ & $\delta(d_3,
    d_2)$ &\\
\hline
    $\rvx_{1c}$ & $x_{1c}\in
    S_{\rvx}$ & $\delta(x_{1c},0)$ &\\
\hline
    $\rvx_{1d}$ & $x_{1d}\in
    S_{\rvx}$ & $\delta(x_{1d},0)$ &\\
\hline
    $\rvx_{3c}$ & $x_{3c}\in S_\rvx$ &
    $\delta(x_{3c}, x_{2c})$ &\\
\hline
    $\rvx_{3d}$ & $x_{3d}\in S_\rvx$ &
    $\delta(x_{3d}, x_{2d})$ &\\
\hline
    $\rvx_{5c}$ & $x_{5c}\in S_\rvx$ &
    $\delta(x_{5c}, x_{4c})$ &\\
\hline
    $\rvx_{5d}$ & $x_{5d}\in S_\rvx$ &
    $\delta(x_{5d}, x_{4d})$ &\\
\EndQBNetTabular
}

$\qbnet$ looks much more complicated
than $\cbnet$, but it really isn't, since
most of its node matrices are
delta functions which quickly disappear
when adding over node states.

According to
Table \ref{tab:scat-qbnet},
the probability amplitude for
the external (aka leaf) nodes is given by

\beq
\begin{array}{l}
A(a_5, b_5, c_3, d_3, x_{5c}, x_{5d})=\\
=
\sum_{\cdot }
\sqrt{
P_{\rva}(\punto{a}_2)
P_{\rvb}(\punto{b}_2)
P_{\rvx|\rva, \rvb}(\punto{x}_{2c}| \punto{a}_{3},
\punto{b}_{3})
P_{\rvc|\rvx}(\punto{c}_{2}| \punto{x}_{3c})
P_{\rvd|\rvx}(\punto{d}_{2}| \punto{x}_{3d})
}\\
\;\;\;\;
\theta(\punto{a}_{2}=\punto{a}_{3}=\punto{a}_{4}=a_5)
\theta(\punto{b}_{2}=\punto{b}_{3}=\punto{b}_{4}=b_5)\\
\;\;\;\;
\theta(\punto{x}_{2c}=\punto{x}_{3c}=\punto{x}_{4c
}=x_{5c})
\theta(\punto{x}_{2d}=\punto{x}_{3d}=\punto{x}_{4d
}=x_{5d})
\theta(x_{5c}= x_{5d})\\
\;\;\;\;
\theta(\punto{c}_{2}=c_3)
\theta(\punto{d}_{2}=d_3)\\
\;\;\;\;
\theta(\punto{a}_{1}=\punto{b}_{1}=\punto{c}_{1}=
\punto{d}_{1}=\punto{x}_{1c}=\punto{x}_{1d}=0)
\end{array}
\;,
\label{eq:scat-leaf-amp-dots}
\eeq
where we have summed over all internal (non-leaf) nodes.
Eq.(\ref{eq:scat-leaf-amp-dots}) immediately reduces to

\beq
\begin{array}{l}
A(a_5, b_5, c_3, d_3, x_{5c}, x_{5d})=\\
=
\sqrt{
P_{\rva}(a_5)
P_{\rvb}(b_5)
P_{\rvx|\rva, \rvb}(x_{5c}| a_{5}, b_{5})
P_{\rvc|\rvx}(c_{3}| x_{5c})
P_{\rvd|\rvx}(d_{3}| x_{5d})
}
\theta(x_{5c}= x_{5d})
\end{array}
\;.
\label{eq:scat-leaf-amp}
\eeq
Eq.(\ref{eq:scat-leaf-amp}) shows that
the net $\qbnet$ that we constructed from
the net $\cbnet$ by following steps 1 and 2
satisfies the definition
Eq.(\ref{eq:q-embed-net-def}) that we
gave earlier for a q-embedding
of $\cbnet$. The probability distribution of
the states of the external nodes
of the QB net $\qbnet$
contains all the probabilistic information of
the original CB net $\cbnet$. Hurray!

From Eq.(\ref{eq:scat-leaf-amp}),
 it is clear that by
running $\qbnet$ on a quantum computer
(or similar quantum system), we can
calculate
any conditional probability that one
would want to calculate for $\cbnet$.
For example,
suppose we wanted to calculate
$P_{\rva, \rvd|\rvx}$.
Run $\qbnet$ on the quantum computer
several times, each time measuring
nodes $\rva_5, \rvd_3$ and $\rvx_{5d}$
and not measuring all other external nodes.
The resulting measurements will be distributed
according to $P_{\rva, \rvd, \rvx}$.
Taking the magnitude squared
of the amplitude and summing
the result over the states of the
un-measured external nodes will
be performed automatically by Nature.
The laws of quantum mechanics guarantee it.
Proceed in the same way to calculate $P_{\rvx}$.
Run $\qbnet$ on the quantum computer
several times, each time measuring
node $\rvx_{5d}$
and not measuring all other external nodes.
Finally divide $P_{\rva, \rvd, \rvx}$ by
$P_{\rvx}$ on a classical (or quantum?) computer.

The q-embedding of a CB net,
as defined by Eq.(\ref{eq:q-embed-net-def}),
is not unique. For example,
we could have defined the net $\qbnet$
given by Fig.\ref{fig:scat-qbnet} without
nodes  $\rva_3$
and $\rvb_3$. We chose to
include such nodes for pedagogical reasons.

To run a QB net on a quantum computer,
we need to replace the QB net by an
equivalent  SEO
that a quantum computer can understand.
This can be done with the help of a
quantum compiler
\cite{Tucci-how-to-compile}\cite{Tucci-qubiter}.
One could compile individually each
node representing a q-embedding, or one
could compile whole
subgraphs of the QB net all at once.
Note that
it may suffice to find
a SEO that is only approximately
(within a certain
precision) equivalent
instead of exactly equivalent to the QB net.
This may be true if,
for example,
the probabilities associated with the CB net
that was q-embedded were not specified
too precisely to
begin with.

Suppose $a_1, a_2, \ldots a_\nu$
belong to a finite set $S_\rva$,
and suppose that they are
distributed according to a probability
distribution $P_\rva$. What number $\nu$
of samples $a_i$ is
necessary to estimate $P_\rva$ within
a given precision? This question is directly
relevant to our method for estimating
probabilities by running a QB net on
a quantum computer.
We will not give a detailed answer
to this question here.
For an answer, the reader can consult any book
on the mathematical theory of Statistics.
An imprecise rule of
thumb is that if the support of $P_\rva$
has $\nu_0$ elements, then
$\nu$ must be at least as large as $\nu_0$;
i.e., one needs at least ``one data
point per bin" to estimate $P_\rva$
with any decent accuracy.

We have given a method for calculating,
via a quantum computer, the conditional
probabilities associated with a CB net.
Does our method have an advantage
in time complexity with respect to classical
methods for calculating the same probabilities?
We will not give a detailed
answer to this question here.
The answer must be yes, sometimes. After all,
our method generalizes the
algorithms by Deutsch-Jozsa,
Simon, Grover, etc., and these
 are known to have a
complexity advantage.

\begin{figure}[h]
    \begin{center}
    \epsfig{file=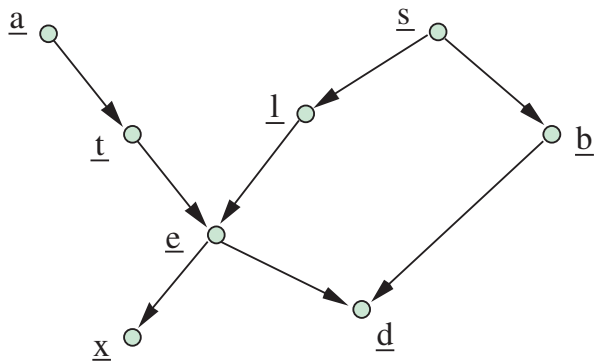, height=2in}
    \caption{CB net for lung disease diagnosis.
We show how to construct a q-embedding for this
CB net.}
    \label{fig:lung-cbnet}
    \end{center}
\end{figure}

To conclude this section, we will
present a second, more complicated
 example of our method of
 finding a q-embedding for a CB net.
A CB net (first given in Ref.\cite{L-S})
for lung disease diagnosis is defined by
Fig.\ref{fig:lung-cbnet} and
Table \ref{tab:lung-cbnet}.

{\footnotesize
\BeginCBNetTabular      \label{tab:lung-cbnet}
    $\rva$ & $a \in Bool$ & $P(\rva=1)=.01$ &
Visited Asia?\\
\hline
    $\rvb$ & $b \in Bool$ & $\begin{array}{l}
    P(\rvb=1 | \rvs=1)=.60 \\
    P(\rvb=1 | \rvs=0)=.30
    \end{array}$ & Bronchitis?\\
\hline
    $\rvd$ & $d \in Bool$ & $\begin{array}{l}
    P(\rvd=1 | \rve=1, \rvb=1)=.90 \\
    P(\rvd=1 | \rve=1, \rvb=0)=.70 \\
    P(\rvd=1 | \rve=0, \rvb=1)=.80 \\
    P(\rvd=1 | \rve=0, \rvb=0)=.10
    \end{array}$ & Dyspnea(trouble breathing)?\\
\hline
    $\rve$ & $e \in Bool$ & $P(e | l,t)=\delta(e,
l \vee t) $ & Either TB or Lung Cancer?\\
\hline
    $\rvl$ & $l \in Bool$ & $\begin{array}{l}
    P(\rvl=1 | \rvs=1)=.10 \\
    P(\rvl=1 | \rvs=0)=.01
    \end{array}$ & Lung Cancer?\\
\hline
    $\rvs$ & $s \in Bool$ & $P(\rvs=1)=.5$ &
Smokes?\\
\hline
    $\rvt$ & $t \in Bool$ & $\begin{array}{l}
    P(\rvt=1 | \rva=1)=.05 \\
    P(\rvt=1 | \rva=0)=.01
    \end{array}$ & Tuberculosis?\\
\hline
    $\rvx$ & $x \in Bool$ & $\begin{array}{l}
    P(\rvx=1 | \rve=1)=.98 \\
    P(\rvx=1 | \rve=0)=.05
    \end{array}$ & Positive X-ray?\\

\EndCBNetTabular
}

\begin{figure}
    \begin{center}
    \epsfig{file=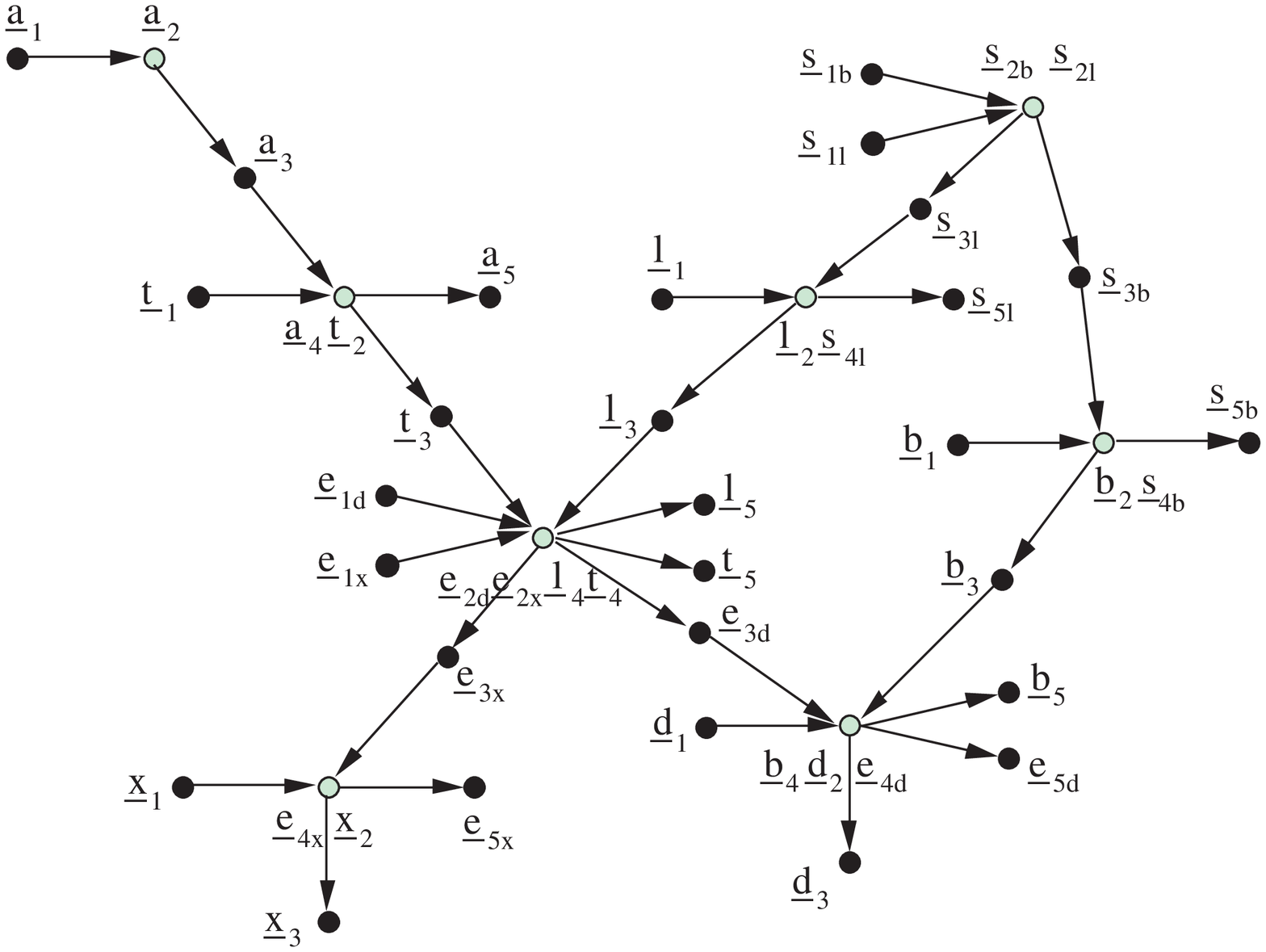, width=5.5in}
    \caption{A QB Net which is a q-embedding
    for the CB net of Fig.\ref{fig:lung-cbnet}.}
    \label{fig:lung-qbnet}
    \end{center}
\end{figure}

If one follows the two steps that were
described earlier in this section,
one obtains the QB net defined
by Fig.\ref{fig:lung-qbnet} and
Table \ref{tab:lung-qbnet}.

{\footnotesize
\BeginQBNetTabularNC      \label{tab:lung-qbnet}
    $\rva_1$ & $a_1 \in Bool$ & $\delta(a_1, 0)$
\\
\hline
    $\rva_2$ & $a_2 \in Bool$ &
$A(a_2|a_1=0)=\sqrt{P_\rva(a_2)}$ \\
\hline
    $\rva_3$ & $a_3 \in Bool$ & $\delta(a_3, a_2)$
\\
\hline
    $(\rva_4, \rvt_2)$ & $(a_4, t_2)\in Bool^2$ &
$A(a_4, t_2| a_3, t_1=0)=\sqrt{P_{\rvt|\rva}(t_2|a_3)} \delta(a_4,
a_3)$\\
\hline
    $\rva_5$ & $a_5\in Bool$ & $\delta(a_5, a_4)$
\\
\hline
    $\rvb_1$ & $b_1 \in Bool$ & $\delta(b_1, 0)$
\\
\hline
    $(\rvb_2, \rvs_{4b})$ & $(b_2, s_{4b})\in
Bool^2$ & $A(b_2, s_{4b}| b_1=0, s_{3b})= \sqrt{P_{\rvb|\rvs}(b_2
| s_{3b})} \delta(s_{4b}, s_{3b})$
\\
\hline
    $\rvb_3$ & $b_3\in Bool$ & $\delta(b_3, b_2)$
\\
\hline
    $(\rvb_4, \rvd_2, \rve_{4d})$ & $(b_4, d_2,
e_{4d}) \in Bool^3$ & $A(b_4, d_2, e_{4d}|b_3, d_1=0, e_{3d})=
\sqrt{P_{\rvd|\rvb,\rve}(d_2 | b_3, e_{3d})} \delta(b_4, b_3)
\delta(e_{4d}, e_{3d})$
\\
\hline
    $\rvb_5$ & $b_5\in Bool$ &
 $\delta(b_5, b_4)$
 \\
\hline
    $\rvd_1$ & $d_1\in Bool$ & $\delta(d_1, 0)$
\\
\hline
    $\rvd_3$ & $d_3\in Bool$ & $\delta(d_3, d_2)$
\\
\hline
    $\rve_{1d}$ & $e_{1d}\in Bool$
& $\delta(e_{1d}, 0)$
\\
\hline
    $\rve_{1x}$ & $e_{1x}\in Bool$
& $\delta(e_{1x}, 0)$
\\
\hline
    $(\rve_{2d}, \rve_{2x}, \rvl_4, \rvt_4)$
& $(e_{2d}, e_{2x}, l_4, t_4)$ & $A(e_{2d}, e_{2x}, l_4, t_4|
e_{1d}=0, e_{1x}=0, l_3, t_3)=$
\\
&$\in Bool^4$ &$\sqrt{P_{\rve|\rvl, \rvt}(e_{2d}|l_3, t_3)}
\delta(e_{2x},e_{2d}) \delta(l_4, l_3)\delta(t_4, t_3)$
\\
\hline
    $\rve_{3d}$ & $e_{3d}\in Bool$ &
$\delta(e_{3d}, e_{2d})$
\\
\hline
    $\rve_{3x}$ & $e_{3x}\in Bool$ &
$\delta(e_{3x}, e_{2x})$
\\
\hline
    $(\rve_{4x}, \rvx_2)$& $(e_{4x}, x_2)\in
Bool^2$ & $A(e_{4x}, x_2| e_{3x}, x_1=0)=
\sqrt{P_{\rvx|\rve}(x_2|e_{3x})} \delta(e_{4x}, e_{3x})$
\\
\hline
    $\rve_{5d}$ & $e_{5d} \in Bool$ &
    $\delta(e_{5d}, e_{4d})$
\\
\hline
    $\rve_{5x}$ & $e_{5x} \in Bool$ &
$\delta(e_{5x}, e_{4x})$
\\
\hline
    $\rvl_1$ & $l_1 \in Bool$ & $\delta(l_1, 0)$
\\
\hline
    $(\rvl_2,\rvs_{4l})$ & $(l_2,s_{4l})\in
Bool^2$ & $A(l_2, s_{4l}| l_1=0, s_{3l})=
\sqrt{P_{\rvl|\rvs}(l_2|s_{3l})} \delta(s_{4l}, s_{3l})$
\\
\hline
    $\rvl_3$ & $l_3\in Bool$ & $\delta(l_3, l_2)$
\\
\hline
    $\rvl_5$ & $l_5\in Bool$ &
$\delta(l_5, l_4)$
\\
\hline
    $\rvs_{1b}$ & $s_{1b}\in Bool$ &
$\delta(s_{1b}, 0)$
\\
\hline
    $\rvs_{1l}$ & $s_{1l}\in Bool$ &
$\delta(s_{1l}, 0)$
\\
\hline
    $(\rvs_{2b}, \rvs_{2l})$ & $(s_{2b},
s_{2l})\in Bool^2$ & $A(s_{2b}, s_{2l}| s_{1b}=0, s_{1l}=0)=
\sqrt{P_{\rvs}(s_{2b})} \delta(s_{2l},s_{2b})$
\\
\hline
    $\rvs_{3b}$ & $s_{3b} \in Bool$ &
$\delta(s_{3b}, e_{2b})$
\\
\hline
    $\rvs_{3l}$ & $s_{3l}\in Bool$ &
$\delta(s_{3l}, s_{2l})$
\\
\hline
    $\rvs_{5b}$ & $s_{5b}\in Bool$ &
$\delta(s_{5b}, s_{4b})$
\\
\hline
    $\rvs_{5l}$ & $s_{5l}\in Bool$ &
$\delta(s_{5l}, s_{4l})$
\\
\hline
    $\rvt_1$ & $t_1 \in Bool$ & $\delta(t_1, 0)$
\\
\hline
    $\rvt_3$ & $t_3\in Bool$ & $\delta(t_3, t_2)$
\\
\hline
    $\rvt_5$ & $t_5\in Bool$ &
$\delta(t_5, t_4)$
\\
\hline
    $\rvx_1$ & $x_1\in Bool$ & $\delta(x_1, 0)$
\\
\hline
    $\rvx_3$ & $x_3\in Bool$ & $\delta(x_3, x_2)$
\\
\EndQBNetTabularNC }

According to Table \ref{tab:lung-qbnet},
the probability amplitude for
the external (aka leaf) nodes is given by

\beq
\begin{array}{l}
A(a_5, b_5, d_3, e_{5d}, e_{5x}, l_5,
s_{5b}, s_{5l},t_5,  x_3)=\\
=\sqrt{
P_{\rva}(a_5)
P_{\rvt|\rva}(t_5| a_5)
P_{\rvb|\rvs}(b_5| s_{5b})
P_{\rvd|\rvb, \rve}(d_3| b_5, e_{5d})
P_{\rve|\rvl, \rvt}(e_{5d}| l_5, t_5)
P_{\rvx|\rve}(x_3| e_{5d})
P_{\rvl|\rvs}(l_5|s_{5l})
P_\rvs(s_{5b})
}\\
\;\;\;\;
\theta(e_{5d}, e_{5x})\theta(s_{5b}, s_{5l})
\end{array}
\;.
\eeq

\section{Voting Net and Grover's Microscope}

In this section we will first
present a CB net, call it
$\cbnet$, that describes voting.
Then we will find a QB net $\qbnet$
that is a
q-embedding of $\cbnet$.
In certain cases,
the probabilities that we
wish to find are too small
to be measurable by
running $\qbnet$ on a quantum
computer. However,
we will show that sometimes
 it is possible to
define a new QB net, call it
$\qbnet'$, that magnifies
and makes measurable the
probabilities that were
unmeasurable using $\qbnet$ alone.
We will refer to $\qbnet'$ as
{\bf Grover's
Microscope for} $\qbnet$, because
$\qbnet'$ is closely
related to Grover's algorithm,
and it magnifies the
probabilities found with $\qbnet$.

\begin{figure}[h]
    \begin{center}
    \epsfig{file=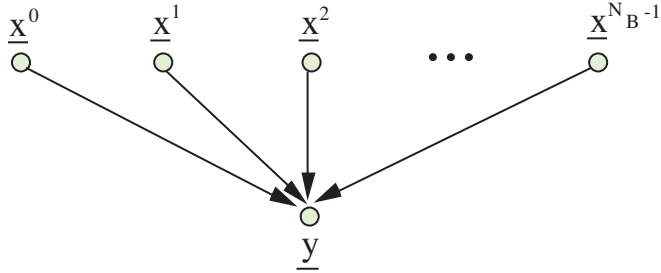, height=1.5in}
    \caption{``Voting" CB net.}
    \label{fig:voting-cbnet}
    \end{center}
\end{figure}

Suppose $y\in Bool$ and $\vec{x} = (x^0, x^1,
\ldots , x^{\nb-1})\in Bool^\nb$.
Consider the
CB net (``voting net")
defined by Fig.\ref{fig:voting-cbnet}
and Table \ref{tab:voting-cbnet}.

\BeginCBNetTabular      \label{tab:voting-cbnet}
    $\rvx^i\;\;\mbox{for all}\;\; i\in Z_{0, \nb-1}$ &
     $x^i\in Bool$
& $P(x^i)$ &\\
\hline
    $\rvy$ & $y\in Bool$ & $P(y|\vec{x})$ & \\
\EndQBNetTabular

Henceforth, we will abbreviate
$P(y=0|\vec{x}) = p_i $ and
$P(y=1|\vec{x}) = q_i $, where
$i=dec(\vec{x})\in Z_{0, \ns-1}$.
Hence $p_i + q_i = 1$ for
all $i\in Z_{0, \ns-1}$.
In general, the probability matrix
$P(y|\vec{x})$
has $2^\nb$ free parameters (namely,
$p_i$ for all $i\in Z_{0, \ns-1}$).
This number of parameters
is a forbiddingly large
for large $\nb$.
To ease the task of specifying $P(y|\vec{x})$ ,
it is common to impose additional  constraints on
$P(y|\vec{x})$.
An interesting special type of $P(y|\vec{x})$
is
{\bf deterministic} $pd(Bool|Bool^\nb)$ {\bf
matrices};
that is, those that can be expressed in the form

\beq
P(y|\vec{x}) = \delta(y , f(\vec{x}))
\;,
\eeq
where $f:Bool^\nb\rarrow Bool$.
In this case, the voting net can be used
to pose the {\bf satisfiability problem (SAT)}:
given $y=0$, find the most likely
 $\vec{x}\in Bool^\nb$; in other words,
 find those $\vec{x}$ for which $f(\vec{x})=0$.
 We say $f$ is {\bf AND-like} if all $p_i$
equal zero except for one
$p_i$ which equals one.
For example, for $\nb=2$,
if $f$ is an AND gate, then

\beq
P(y|\vec{x})_{AND}=\left\{
\begin{array}{r}
(x^0, x^1) \rarrow\\
\begin{array}{llllll}
&&\vline 00 &01 & 10 & 11\\
\hline
y\downarrow
&0&\vline 1& 1& 1 &0\\
&1&\vline 0& 0& 0 &1\\
\end{array}
\end{array}
\right.
\;.
\eeq
A slightly more general type of $P(y|\vec{x})$
is
{\bf quasi-deterministic} $pd(Bool|Bool^\nb)$ {\bf
matrices};
that is, those that can be expressed in the form

\beq
P(y|\vec{x}) =
\sum_{\vec{t}}
\delta(y, f(\vec{t}))
P(t^0|x^0)P(t^1|x^1)\ldots P(t^{\nb-1}|x^{\nb-1})
\;,
\eeq
where
$f:Bool^\nb\rarrow Bool$
and we sum over all
$\vec{t} = (t^0, t^1, \ldots , t^{\nb-1})\in
Bool^\nb$.
When $f(\vec{t})= t^0\vee t^1 \vee \ldots
\vee t^{\nb-1}$,
$P(y|\vec{x})$ is called a {\bf noisy-OR}.
Appendix \ref{app:det-mat}
discusses how to q-embed deterministic
 $pd(Bool|Bool^\nb)$ matrices, and
how to express such q-embeddings as a SEO .
Appendix \ref{app:quasi-det-mat}
discusses the same thing for quasi-deterministic
$pd(Bool|Bool^\nb)$ matrices.

\begin{figure}[h]
    \begin{center}
    \epsfig{file=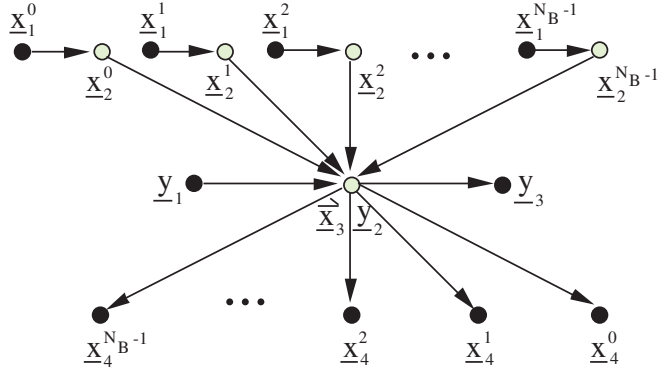, height=2in}
    \caption{A QB Net which is a q-embedding
    for the CB net of Fig.\ref{fig:voting-cbnet}.}
    \label{fig:voting-qbnet}
    \end{center}
\end{figure}

A q-embedding for the
CB net defined by Fig.\ref{fig:voting-cbnet}
and Table \ref{tab:voting-cbnet}
is given by the QB net defined by
Fig.\ref{fig:voting-qbnet}
and Table \ref{tab:voting-qbnet}.

{\footnotesize
\BeginQBNetTabular      \label{tab:voting-qbnet}
    $\vec{\rvx}_1$ & $\vec{x}_1\in Bool^\nb$ &
$\delta(\vec{x}_1, 0)$ &\\
\hline
    $\vec{\rvx}_2$& $\vec{x}_2\in Bool^\nb$&
$A(\vec{x}_2|\vec{x}_1=0)=\sqrt{P_{\vec{\rvx}}
(\vec{x}_2)}$ &\\
\hline
    $(\vec{\rvx}_3, \rvy_2)$& $(\vec{x}_3, y_2)\in
Bool^{\nb+1}$ &
$A(\vec{x}_3, y_2|\vec{x}_2, y_1=0)=
\sqrt{P_{\rvy|\vec{\rvx}}(y_2|\vec{x}_2)}\delta(
\vec{x}_3, \vec{x}_2)$  &\\
\hline
    $\vec{\rvx}_4$& $\vec{x}_4\in Bool^\nb$ &
$\delta(\vec{x}_4, \vec{x}_3)$ &\\
\hline
    $\rvy_1$& $y_1\in Bool$ & $\delta(y_1, 0)$ &\\
\hline
    $\rvy_3$& $y_3\in Bool$ & $\delta(y_3, y_2)$ &\\
\EndQBNetTabular
}

According to Table \ref{tab:voting-qbnet},
the probability amplitude for the leaf (external)
nodes is

\begin{subequations}
\label{eq:voting-ampl}

\begin{eqnarray}
\lefteqn{A(\vec{x}_4, y_3)=}\nonumber\\
&=&\sum_{\cdot }
\sqrt{
P_{\vec{\rvx}}(\vec{\punto{x}}_2)
P_{\rvy|\vec{\rvx}}(\punto{y}_{2}|
\vec{\punto{x}}_2)
}
\theta(\punto{y}_2=y_3)
\theta(\vec{\punto{x}}_2=
\vec{\punto{x}}_3=\vec{x}_4)
\theta(\vec{\punto{x}}_1 = \punto{y}_1 =0)
\\
&=&
\sqrt{
P_{\vec{\rvx}}(\vec{x}_4)
P_{\rvy|\vec{\rvx}}(y_3| \vec{x}_4)
}
\;.
\end{eqnarray}
\end{subequations}

To fully specify the QB net for voting, we
need to extend $A(\vec{x}_2| \vec{x}_1=0)$ and
$A(\vec{x}_3, y_2| \vec{x}_2, y_1=0)$ into unitary
matrices by adding columns to them. This can
always be accomplished by applying the Gram-Schmidt
algorithm.
But sometimes one can guess a matrix extension
and applying Gram-Schmidt becomes
unnecessary.
If $P_{\vec{\rvx}}$ is uniform (i.e.,
$P(\vec{x}) = 1/\ns$ for all $\vec{x}$,
 which means
there is no a priori information
about $\vec{x}$), then
$A(\vec{x}_2| \vec{x}_1=0) = 1/\sqrt{\ns}$.
In this case, we can
 extend  $A(\vec{x}_2| \vec{x}_1=0)$
into the
unitary matrix

\beq
[A(\vec{x}_2| \vec{x}_1)] = \hat{H}_\nb
\;.
\label{eq:voting-mat-exten1}
\eeq
(This works because all
 entries of the first column of
$\hat{H}_\nb$ are
equal to $1/\sqrt{\ns}$.)
As to extending $A(\vec{x}_3, y_2| \vec{x}_2,
y_1=0)$,
this can be done as follows.
Define

\beq
\Delta_p = diag(\sqrt{p_0}, \sqrt{p_1}, \ldots ,
\sqrt{p_{\ns-1}})
\;,
\eeq
and

\beq
\Delta_q = diag(\sqrt{q_0}, \sqrt{q_1}, \ldots ,
\sqrt{q_{\ns-1}})
\;.
\eeq
A possible way of extending $A(\vec{x}_3, y_2|
\vec{x}_2, y_1=0)$
into a unitary matrix is

\beq
[A(\vec{x}_3, y_2| \vec{x}_2, y_1)]=
\left(
\begin{array}{cc}
\Delta_p & - \Delta_q\\
\Delta_q & \Delta_p
\end{array}
\right)
\;.
\label{eq:voting-mat-exten2}
\eeq
Unitary matrices of this kind are called
D-matrices
in Ref.\cite{Tucci-qubiter}.
Ref.\cite{Tucci-qubiter} shows how to
decompose any D-matrix into a SEO.

Earlier, we explained how to estimate a
conditional probability
for a CB net
by running
a QB net $\nu$ times on a quantum computer.
If we wanted to find $P(y|x^0, x^1)$
for the voting CB net,
then the number of runs
$\nu$
required to estimate $P(y|x^0, x^1)$
with moderate accuracy would not be too onerous,
because
the domain of $P(y|x^0, x^1)$ is
$Bool^3$, which contains
only 8 points. But what if we wanted to
estimate $P(y|\vec{x})$?
For large $\nb$, the domain of
$P(y|\vec{x})$ is very large ($2^{\nb+1}$ points).
If the support of $P(y|\vec{x})$
occupies a large fraction of this domain,
then the number of runs $\nu$ required to estimate
$P(y|\vec{x})$ with moderate accuracy
 is forbiddingly large.
However, there are some cases in which ``Grover's
Microscope" can come to the rescue, by
allowing us to amplify certain
salient features of
$P(y|\vec{x})$
so that they become measurable
in only a few runs.

Next we will discuss Grover's Microscope
for the voting QB net
defined
by Fig.\ref{fig:voting-qbnet} and
Table \ref{tab:voting-qbnet}. For simplicity,
we will
assume that $P_{\vec{\rvx}}$ is uniform.

Let $\vec{\kappa}=(\kappa_0, \kappa_1, \ldots,
 \kappa_{\nb-1})$
label  $\nb$ bits
and let $\tau$ label
another bit. Assume that $\tau$ and
 all the $\kappa_i$ are distinct.
Define

\beq
\ket{\phi_p}_{\vec{\kappa}}=
\phi_p = (\sqrt{p_0}, \sqrt{p_1}, \ldots ,
\sqrt{p_{\ns-1}})^T
\;,
\eeq

\beq
\ket{\phi_q}_{\vec{\kappa}}=
\phi_q = (\sqrt{q_0}, \sqrt{q_1}, \ldots ,
\sqrt{q_{\ns-1}})^T
\;,
\eeq
and

\beqa
\ket{\Psi}
&=&
\frac{1}{\sqrt{\ns}}\left(
\ket{0}_{\tau}\ket{\phi_p}_{\vec{\kappa}} +
\ket{1}_{\tau}\ket{\phi_q}_{\vec{\kappa}}
\right)
\nonumber\\
&=&
\frac{1}{\sqrt{\ns}}
\left[
\left(\begin{array}{c}1\\0\end{array}\right)
\otimes \phi_p
+
\left(\begin{array}{c}0\\1\end{array}\right)
\otimes \phi_q
\right]
= \frac{1}{\sqrt{\ns}}
\left(
\begin{array}{c}
\phi_p\\
\phi_q
\end{array}
\right)
= \Psi
\;.
\eeqa
Since $p_i + q_i = 1$ for all $i$,
$\phi_p^T\phi_p + \phi_q^T\phi_q = \ns$.
According to Eq.(\ref{eq:voting-ampl}), when
$P_{\vec{\rvx}}$ is uniform,
the voting QB net
fully specifies a unitary matrix $U_{net}$
such that

\beq
\ket{\Psi} = U_{net} \ket{0}_{\vec{\kappa}}
\ket{0}_\tau
\;.
\eeq

Define orthonormal vectors $e_0$ and $e_1$ by

\beq
e_0 = \left(
\begin{array}{c}
\hat{\phi}_p\\
0
\end{array}
\right)
\;,\;\;
e_1 = \left(
\begin{array}{c}
0\\
\hat{\phi}_q
\end{array}
\right)
\;,
\eeq
where $\hat{V}$ is a unit vector
in the direction of $\vec{V}$.
If $P(y|\vec{x})$ is deterministic
with AND-like $f$,
then  all components of $e_0$ are
 zero except for the one at
the target state $j_{targ}$.

In terms of $e_0, e_1$, $\Psi$
can be expressed as

\beq
\Psi = \frac{1}{\sqrt{\ns}}
\left(
\begin{array}{c}
\phi_p\\
\phi_q
\end{array}
\right)=
\frac{1}{\sqrt{\ns}}
(  |\phi_p|e_0
+ |\phi_q|e_1)
\;.
\eeq
It is convenient to define a vector $\Psi_\perp$
orthogonal to $\Psi$:

\beq
\Psi_\perp =
\frac{1}{\sqrt{\ns}}
(  |\phi_q|e_0
- |\phi_p|e_1)
\;.
\eeq
If
$P(y|\vec{x})$ is deterministic
with AND-like $f$, then
$|\phi_p|=1$ and $|\phi_q| = \sqrt{\ns-1}$ so,
for large $\ns$, $\Psi\approx e_1$
and
$\Psi_\perp\approx e_0$.
For an arbitrary angle $\alpha$, let

\beq
\Psi'_\perp= \frac{1}{\sqrt{\ns}}
\left[
(c_{\frac{\alpha}{2}} |\phi_q|
+ s_{\frac{\alpha}{2}}|\phi_p|)e_0 +
(s_{\frac{\alpha}{2}} |\phi_q|
- c_{\frac{\alpha}{2}}|\phi_p|)e_1
\right]
\;,
\eeq
where $s_A = \sin A$ and
$c_A = \cos A$ for any angle $A$.
Let $\angle(x,y)$ denote
 the angle between 2 vectors
$x$ and $y$.
Note that
$\angle(\Psi'_\perp, \Psi_\perp) =\alpha/2$.
We define
$\angle(e_1, \Psi) =\theta/2$.

\begin{figure}[h]
    \begin{center}
    \epsfig{file=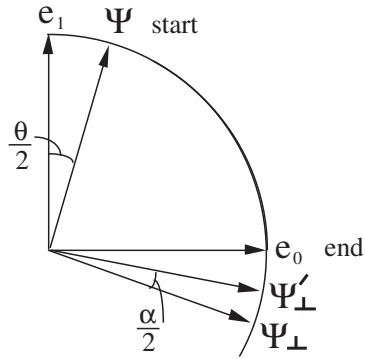, height=2in}
    \caption{Various vectors relevant to
Grover's Microscope.}
    \label{fig:Grover-micro-rot}
    \end{center}
\end{figure}

Fig.\ref{fig:Grover-micro-rot} portrays various vectors that arise
in explaining Grover's Microscope. Note that $\Psi'_\perp = e_0$
when $\alpha=\theta$.

Since we plan to stay within
the two dimensional vector space
with orthonormal basis $e_0, e_1$,
it is convenient to switch matrix representations.
Within $span(e_0, e_1)$,
$e_0, e_1$ can be represented more simply by:

\beq
e_0 =
\left(
\begin{array}{c}
1\\0
\end{array}
\right)
\;,\;\;
e_1 =
\left(
\begin{array}{c}
0\\1
\end{array}
\right)
\;.
\eeq
If
$e_0, e_1$ are represented in this way, then

\beq
\Psi = \frac{1}{\sqrt{\ns}}
\left(
\begin{array}{c}
|\phi_p|\\
|\phi_q|
\end{array}
\right)
\;,
\eeq

\beq
\Psi_\perp = \frac{1}{\sqrt{\ns}}
\left(
\begin{array}{c}
|\phi_q|\\
-|\phi_p|
\end{array}
\right)
\;,
\eeq
and

\beq
\Psi'_\perp = W\Psi
\;,\;\;\mbox{where}\;\;
W=
\left(
\begin{array}{cc}
c_{\frac{\alpha}{2}} & -s_{\frac{\alpha}{2}}\\
s_{\frac{\alpha}{2}} & c_{\frac{\alpha}{2}}
\end{array}
\right)
\left(
\begin{array}{cc}
0 & 1\\
-1 & 0
\end{array}
\right)
\;.
\eeq
The matrix $\left(\begin{array}{cc}
0 & 1\\
-1 & 0
\end{array}\right)$ is a clockwise rotation by
$\pi/2$ in space $span(e_0, e_1)$.
Thus, $W$ equals a
clockwise rotation by $\pi/2$ followed
by a counter-clockwise rotation by
$\alpha/2$.

Define the following reflection operators

\beq
R_0 = 1 - 2\Pi_{\ket{0}_{\vec{\kappa}}}\Pi_{\ket{0}_\tau}
=
(-1)^{\Pi_{\ket{0}_{\vec{\kappa}}}
\Pi_{\ket{0}_\tau}}
\;,
\eeq

\beq
R_\Psi = U_{net} R_0 U^\dagger_{net}
\;,
\eeq

\beq
R_{\Psi'_\perp} = W R_\Psi W^\dagger = W U_{net}
R_0 U^\dagger_{net} W^\dagger
\;.
\eeq
From Eq.(\ref{eq:double-reflection-gen}),
it follows that

\beq
-R_\Psi R_{\Psi'_\perp} =
c_{\alpha}\Psi\Psi^T
-s_{\alpha}\Psi\Psi^T_\perp
+s_{\alpha}\Psi_\perp\Psi^T
+c_{\alpha}\Psi_\perp\Psi^T_\perp
\;.
\eeq
Thus,
$-R_\Psi R_{\Psi'_\perp}$ rotates vectors
in $span(e_0, e_1)$, clockwise
 by an angle $\alpha$.

Grover's Microscope can be
summarized by the following
equation

\beq
(-R_\Psi R_{\Psi'_\perp})^r
\Psi \approx e_0
\;,
\eeq
for some integer $r$ to be determined,
where ``$\approx$" means approximation at large $\ns$.
What this means is that our system starts in state
$\Psi$ and is rotated consecutively $r$ times,
each time by a small angle
$\alpha$, until it arrives at
the state  $e_0$. If
$P(y|\vec{x})$ is deterministic
with AND-like $f$, then
measuring state $e_0$
 yields the target state $j_{targ}$.

The optimum number $r$ of iterations is

\beq
r\alpha \approx \frac{\pi}{2}(1 + 2k)
\;
\eeq
for some integer $k$.
Note that $\cos (\theta/2) =
 \av{\Psi| e_1} = |\phi_q|/\sqrt{\ns}$ so,
in general, $\theta$ depends on $|\phi_p|$ (or on
$|\phi_q|=\sqrt{\ns-|\phi_p|^2}$). If $P(y|\vec{x})$ is
deterministic with AND-like $f$, then $|\phi_p|=1$ and $|\phi_q| =
\sqrt{\ns-1}$. In this case, it is convenient to choose
$\alpha=\theta$, so that $\Psi_\perp' = e_1$ and
Figs.\ref{fig:Grover-rot} and \ref{fig:Grover-micro-rot} become
the same diagram
 under the mapping
 $\Psi\rarrow \mu$
 and $\Psi'_\perp \rarrow \phi=e_0$.
 Then the optimum number $r$ of iterations
 for Grover's original algorithm and
 for Grover's Microscope are equal.
 If we don't know ahead of time
 the value of $|\phi_p|$, then
 setting $\theta= \alpha$ will make both
 $r$ and $\alpha$ depend on
 the unknown $|\phi_p|$,
 although the product $r\alpha$ will
 still be independent of it.

Let

\beqa
U_{\mu scope} &=&
\left(\begin{array}{cc}
0 & 1\\
-1 & 0
\end{array}\right)\nonumber\\
&=&-e_1 e_0^T + e_0 e_1^T\nonumber \\
&=&
-\Psi \Psi_\perp^T + \Psi_\perp \Psi^T
\;.
\eeqa
Note that

\beq
U_{\mu scope}\Psi=
\Psi_\perp
\;.
\eeq
From the point of view of quantum compiling,
Grover's Microscope approximates
the $\pi/2$
rotation $U_{\mu scope}$
by
the $r$-fold product of $-R_\Psi R_{\Psi'_\perp}$,
where we assume that
$-R_\Psi R_{\Psi'_\perp}$
 can be shown to have a
SEO of low
(polynomial in $\nb$) complexity.
(If such a low complexity SEO cannot be found,
then it is pointless to divide
$U_{\mu scope}$ into $r$ iterations of
$-R_\Psi R_{\Psi'_\perp}$, and we
might be better off
compiling $U_{\mu scope}$ all at once.)

\begin{appendix}

\section{Appendix: Deterministic
$pd(Bool | Bool^\nb)$ matrices} \label{app:det-mat}

In this Appendix, we will first define a
special kind of probability matrices which
we call deterministic $pd(Bool | Bool^\nb)$
 matrices.
Then we will show how
such probability matrices
can be q-embedded, and how their
q-embedding can be expressed as a SEO.

Suppose $y\in Bool$ and $\vec{x} = (x^0, x^1,
\ldots , x^{\nb-1})\in Bool^\nb$.
Let  $f:Bool^\nb\rarrow Bool$.

We will say that  $f$ is {\bf AND-like} if
$f(\vec{x}) = \theta(\vec{x}=\vec{x}_{targ})$
for some target vector $\vec{x}_{targ}\in
Bool^\nb$.
An AND-like $f$ maps all $\vec{x}$ into zero
except for  $\vec{x}_{targ}$ which it maps into
one. Thus, $|f^{-1}(1)|=1$.
An example of an AND-like $f$ is the multiple
AND gate $f(\vec{x})= x^0\wedge x^1 \wedge \ldots
\wedge x^{\nb-1}$,
which can also be expressed as $f(\vec{x})=
\theta[\vec{x} = (1, 1, \ldots, 1)]$.

We will say that  $f$ is {\bf OR-like} if
$f(\vec{x}) = \theta(\vec{x}\neq\vec{x}_{targ})$
for some target vector $\vec{x}_{targ}\in
Bool^\nb$.
An OR-like $f$ maps all $\vec{x}$ into one
except for  $\vec{x}_{targ}$ which it maps into
zero. Thus, $|f^{-1}(0)|=1$.
An example of an OR-like $f$ is the multiple
OR gate $f(\vec{x})= x^0\vee x^1 \vee \ldots \vee
x^{\nb-1}$,
which can also be expressed as $f(\vec{x})=
\theta[\vec{x} \neq (0, 0, \ldots, 0)]$.

We will say that  $f$  has a {\bf single target}
if it is either AND-like or OR-like. If $f$ has
more than
one target (i.e., if $|f^{-1}(0)|$ and
$|f^{-1}(1)|$ are both greater than one),
then we will say that $f$ has {\bf multiple
targets}.

Suppose $y\in Bool$ and $\vec{x} = (x^0, x^1,
\ldots , x^{\nb-1})\in Bool^\nb$.
Let  $f:Bool^\nb\rarrow Bool$.
In this section, we consider
{\bf deterministic} $pd(Bool|Bool^\nb)$ {\bf
matrices};
that is, probability matrices of the form
$P(y|\vec{x}) = \delta(y , f(\vec{x}))$. First let
us consider the case that
$f$ has a single target. For example, for $\nb=2$,
if $f$ is an AND gate

\beq
P(y|\vec{x})_{AND}=\left\{
\begin{array}{r}
(x^0, x^1) \rarrow\\
\begin{array}{llllll}
&&\vline 00 &01 & 10 & 11\\
\hline
y\downarrow
&0&\vline 1& 1& 1 &0\\
&1&\vline 0& 0& 0 &1\\
\end{array}
\end{array}
\right.
\;,
\eeq
and
if $f$ is an
OR gate

\beq
P(y|\vec{x})_{OR}=
\left\{
\begin{array}{r}
(x^0, x^1) \rarrow\\
\begin{array}{llllll}
&&\vline 00 &01 & 10 & 11\\
\hline
y\downarrow
&0&\vline 1& 0& 0 &0\\
&1&\vline 0& 1& 1 &1\\
\end{array}
\end{array}
\right.
\;.
\eeq

Suppose bit value $y$ is stored in the bit
labelled $\tau$.
And suppose bit values
 $x^0, x^1, \ldots, x^{\nb-1}$
are stored in the
bits labelled $\vec{\kappa} = (\kappa_0, \kappa_1,
 \ldots, \kappa_{\nb-1})$.
Define $e_j$ for all $j\in Z_{0, \ns-1}$
to be the $\ns$ dimensional column vector with
$j$th component equal to one and all other
components equal to zero.
Let
$\Pi_j = e_j e_j^T$
and
$\Pi_{targ} = \Pi_{j_{targ}}$, where
 $j_{targ}\in Z_{0, \ns-1} $ is the target state.
$ \Pi_{targ}$ can expressed as product of number
operators.
Indeed, if

\beq
j_{targ} =
\sum_{i=0}^{\nb-1} x_{targ,
i} 2^i
\;,
\eeq
then

\beq
\Pi_{targ}  = \Pi_{j_{targ}}=
\prod_{i=0}^{\nb-1} \left[
\nop(\kappa_i)\theta(x_{targ, i}=1)
+  \nbar(\kappa_i)\theta(x_{targ, i}=0) \right]
\;.
\label{eq:single-targ-pi}
\eeq
For example, if $j_{targ}=0$ then $\Pi_{targ}
=\nbar(\kappa_0)\nbar(\kappa_1)
\ldots\nbar(\kappa_{\nb-1})$.

An AND-like probability matrix
$P(y|\vec{x})$ is
q-embedded within the unitary matrix

\beq
U_{AND-like}=[A(y,\vec{\tilde{x}}|\tilde{y},
\vec{x})]=
\begin{array}{lll}
&\vline \tilde{y}=0 &\tilde{y}=1\\
\hline
y=0&\vline 1-\Pi_{targ}& -\Pi_{targ}\\
y=1&\vline \Pi_{targ}& 1-\Pi_{targ}
\end{array}
\;.
\eeq
Note that

\begin{subequations}
\beqa
U_{AND-like} &=&
 1 + \left(
\begin{array}{cc}-1&-1\\1&-1\end{array}
\right)\otimes\Pi_{targ} \\
&=& 1 + \Pi_{targ}(\vec{\kappa})(-i
\sigma_y(\tau) - 1)\\
&=& [-i
\sigma_y(\tau)]^{\Pi_{targ}(\vec{\kappa})}
\label{eq:multi-cnot}
\;.
\eeqa
\end{subequations}
Eqs.(\ref{eq:single-targ-pi})
and (\ref{eq:multi-cnot}) show how to express
$U_{AND-like}$ as a
qubit rotation with multiple control qubits.
Operations of this kind
can be decomposed into a SEO using
the techniques of
Refs.\cite{Barenco} and \cite{Tucci-qubiter}.

An OR-like probability matrix
$P(y|\vec{x})$ is
q-embedded within the unitary matrix

\beq
U_{OR-like}=[A(y,\vec{\tilde{x}}|\tilde{y},
\vec{x})]=
\begin{array}{lll}
&\vline \tilde{y}=0 &\tilde{y}=1\\
\hline
y=0&\vline \Pi_{targ}& 1-\Pi_{targ}\\
y=1&\vline 1-\Pi_{targ}& -\Pi_{targ}
\end{array}
\;.
\eeq
Note that

\begin{subequations}
\beqa
U_{OR-like} &=&
\left(
\begin{array}{cc}
0 & I_\ns\\I_\ns & 0
\end{array}
\right)
\left(
\begin{array}{cc}
1-\Pi_{targ} & -\Pi_{targ}\\
\Pi_{targ} & 1-\Pi_{targ}
\end{array}
\right)
\\
&=& \sigma_x(\tau)[-i
\sigma_y(\tau)]^{\Pi_{targ}(\vec{\kappa})}
\;.
\eeqa
\end{subequations}

Finally,  let us consider the case when
$f:Bool^\nb\rarrow Bool$
has multiple targets. Let
 $T\subset Z_{0, \ns-1}$ be the set of these targets;
 i.e., either $T=f^{-1}(0)$ or $T=f^{-1}(1)$.
 Define $\Pi_{targ}$ by

\beq
\Pi_{targ} =
\sum_{j\in T}
 \Pi_j
\;.
\label{eq:multi-targ-pi}
\eeq
$\Pi_{targ}$ can be expressed as a product
of number operators. Indeed, each $\Pi_j$
on the right hand side of
Eq.(\ref{eq:multi-targ-pi}) can be
separately expressed,
using Eq.(\ref{eq:single-targ-pi}), as
a product of number operators.
If $T=f^{-1}(1)$, then  $P(y|\vec{x})$
is q-embedded within the unitary matrix

\beqa
U_{multi-targ}&=&[A(y,\vec{\tilde{x}}|\tilde{y},
\vec{x})]=
\begin{array}{lll}
&\vline \tilde{y}=0 &\tilde{y}=1\\
\hline
y=0&\vline 1-\Pi_{targ}& -\Pi_{targ}\\
y=1&\vline \Pi_{targ}& 1-\Pi_{targ}
\end{array}\nonumber\\
&=&
[-i\sigma_y(\tau)]^{\Pi_{targ}(\vec{\kappa})}
\;.
\eeqa

\section{Appendix:
Quasi-deterministic
$pd(Bool | Bool^\nb)$ matrices}
\label{app:quasi-det-mat}

In this Appendix, we will first define a
special kind of probability matrices which
we call quasi-deterministic
$pd(Bool | Bool^\nb)$ matrices.
Then we will show how
such probability matrices
can be q-embedded, and how their
q-embedding can be expressed as a SEO.

\begin{figure}[h]
    \begin{center}
    \epsfig{file=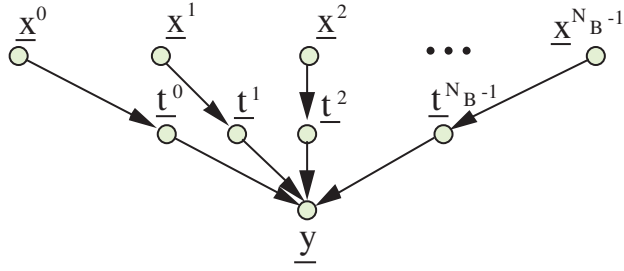, height=1.5in}
    \caption{Quasi-deterministic (``noisy")
    $pd(Bool|Bool^\nb)$ gate.}
    \label{fig:noisy-gate}
    \end{center}
\end{figure}

Suppose $y\in Bool$ and $\vec{x} = (x^0, x^1,
\ldots , x^{\nb-1})\in Bool^\nb$.
Let  $f:Bool^\nb\rarrow Bool$.
In the previous appendix, we considered
deterministic $pd(Bool|Bool^\nb)$ matrices;
that is, probability matrices of the form
$P(y|\vec{x}) = \delta(y , f(\vec{x}))$.
In this section, we will consider
{\bf quasi-deterministic} $pd(Bool|Bool^\nb)$ {\bf
matrices};
that is, probability matrices of the form

\beq
P(y|\vec{x}) =
\sum_{\vec{t}}
\delta(y, f(\vec{t}))
P(t^0|x^0)P(t^1|x^1)\ldots P(t^{\nb-1}|x^{\nb-1})
\;,
\label{eq:noisy-gate}
\eeq
where we sum over all
$\vec{t} = (t^0, t^1, \ldots , t^{\nb-1})\in
Bool^\nb$.
Fig.\ref{fig:noisy-gate} shows a
CB net representation of Eq.(\ref{eq:noisy-gate}).
Examples of quasi-deterministic
$pd(Bool|Bool^\nb)$
matrices are:
 (1)the {\bf noisy OR},
for which $f(\vec{t}) = t^0\vee t^1 \vee\ldots
\vee t^{\nb-1}$;
(2)the {\bf noisy AND},
for which $f(\vec{t}) = t^0\wedge t^1 \wedge\ldots
\wedge t^{\nb-1}$;
(3)the {\bf noisy CNOT},
for which $f(\vec{t}) = t^0\oplus t^1 \oplus
\ldots \oplus t^{\nb-1}$, etc.

For each $\alpha\in Z_{0, \nb-1}$, the
probabilities
$P(\rvt^\alpha=t|\rvx^\alpha=x)$
will be abbreviated by $p_{t,x}^\alpha$ for
$t,x\in Bool$.
$P(\rvt^\alpha=t|\rvx^\alpha=x)$ has two
independent parameters which we may take to be
$p_{01}^\alpha$ (the probability of false
negatives) and
$p_{10}^\alpha$ (the probability of false
positives).
$p_{00}^\alpha$ and $p_{11}^\alpha$ can be
expressed in terms of these
independent parameters:
$p_{00}^\alpha= 1- p_{10}^\alpha$,
$p_{11}^\alpha=1-p_{01}^\alpha$.
Whereas a completely general
probability matrix $P(y|\vec{x})\in
pd(Bool|Bool^\nb)$
has $2^\nb$ free parameters, a quasi-deterministic
$P(y|\vec{x})$ has $2\nb$ free parameters.

Rather than q-embedding the probability matrix
$P(y|\vec{x})$ as a whole, it is convenient
to q-embed separately the probability matrices
$P(y|\vec{t})$ and $P(t^\alpha | x^\alpha)$
for every $\alpha\in Z_{0, \nb-1}$.
$P(y|\vec{t})=\delta(y, f(\vec{t}))$ is a
deterministic $pd(Bool|Bool^\nb)$ matrix so
its q-embedding is discussed in
Appendix \ref{app:det-mat}. As for
$P(t^\alpha | x^\alpha)$, it can be
easily q-embedded as follows. For each
$\alpha \in Z_{0, \nb-1}$, let

\beq
\Delta^\alpha_p =
\left(
\begin{array}{cc}
\sqrt{p_{00}^\alpha} & 0\\
0& \sqrt{p_{01}^\alpha}
\end{array}
\right)
\;,\;\;
\Delta^\alpha_q =
\left(
\begin{array}{cc}
\sqrt{p_{10}^\alpha} & 0\\
0& \sqrt{p_{11}^\alpha}
\end{array}
\right)
\;.
\eeq
$P(t^\alpha | x^\alpha)$ is q-embedded
within the unitary matrix:

\beq
[A(t^\alpha, \tilde{x}^\alpha|
 \tilde{t}^\alpha, x^\alpha) ]=
\left(
\begin{array}{cc}
\Delta^\alpha_p & -\Delta^\alpha_q\\
\Delta^\alpha_q& \Delta^\alpha_p
\end{array}
\right)
\;.
\eeq
Unitary matrices of this kind are called
D-matrices
in Ref.\cite{Tucci-qubiter}.
Ref.\cite{Tucci-qubiter} shows how to
decompose any D-matrix into a SEO.

\end{appendix}


\begin{thebibliography}{99}

\bibitem{Tucci-review}
R.R. Tucci, ``Quantum Information Theory - A
Quantum Bayesian Nets Perspective", ArXiv
eprint quant-ph/9909039 .

\bibitem{D-J}
D. Deutsch and R. Jozsa, Proc. Roy. Soc. of London
A (1992) {\bf 439}, 553.
R. Jozsa, ArXiv eprint quant-ph/9707033 .



\bibitem{Simon}D.R. Simon, {\it Proceedings of the
35th Annual IEEE Symp. on the Found. of Comp.
Sci.} (IEEE Computer Society, Los Alamitos, 1994).
Extended Abstract on page 116. Full Version of the
paper in S.I.A.M. Jour. on Computing, {\bf 26},
Oct 97.

\bibitem{B-V}
E. Bernstein, U. Vazirani,
Proceedings of the 25th Annual ACM Synposium
on Theory of Computing, pages 11-20 (1993)

\bibitem{Grover}
Lov K. Grover, ArXiv eprint quant-ph/9605043

\bibitem{F-T}
T. Toffoli, {\it Automata, Languages and
Programming, 7th Coll.} (Springer Verlag, 1980)
pg. 632.
E. Fredkin, T. Toffoli, Int. Jour. of Th. Phys.
(1982) {\bf 21}, 219.

\bibitem{Barenco}
Barenco et al.
``Elementary gates for quantum computation",
ArXiv eprint quant-ph/9503016


\bibitem{Tucci-qubiter}
R.R. Tucci, ``A Rudimentary Quantum Compiler(2cnd
ed.)", ArXiv eprint quant-ph/9902062 .


\bibitem{Tucci-how-to-compile}
R.R. Tucci, ``How to Compile
a Quantum Bayesian Net",
ArXiv eprint quant-ph/9805016


\bibitem{Noble}B. Noble and J.W. Daniels,
{\it Applied Linear Algebra}, Third Edition
(Prentice Hall, 1988).

\bibitem{QR}
Grover's algorithm expresses an orthogonal
matrix as a product
of real reflections. This is related to
 the QR
decomposition of
Linear Algebra\cite{Noble}, wherein any
real (ditto, complex) matrix
$A$ is expressed as $QR$,
where $Q$ is a product of real (ditto, complex)
``Householder" reflections and $R$
is an upper triangular real
 (ditto, complex) matrix.
A byproduct of the $QR$
decomposition is a method for expanding
an orthogonal (ditto, unitary) matrix as
a product of real (ditto, complex)
Householder reflections.

\bibitem{amplit-amplif}
G. Brassard , P. Hoyer , M. Mosca , A. Tapp,
ArXiv eprint quant-ph/0005055

\bibitem{Younes}
Ahmed Younes, Jon Rowe, Julian Miller,
ArXiv eprint quant-ph/0312022

\bibitem{Shor}P. Shor, {\it Proceedings of the
35th Annual IEEE Symp. on the Found. of Comp.
Sci.} (IEEE Computer Society, Los Alamitos, 1994),
page 124.

\bibitem{Tele}
C.H. Bennett, G. Brassard, C. Cr\'{e}peau, R.
Jozsa, A. Peres, W. Wootters,
Phys. Rev. Lett., 70, 1895 (1993).


\bibitem{real-extension}
Note that the matrix $A$ defined by
Eq.(\ref{eq:gen-mat-q-embed})
will have real entries if the $\xi^{(x)}$ basis is
chosen
to lie in the real $N_\rvx$ dimensional vector
space
and the Gram-Schmidt process is carried out in
that same space.
Thus, one can always find a
q-embedding $A$ for a probability matrix
such that
$A$ is not merely
unitary, but also orthogonal.
However, if $A$ is
destined to become a node matrix
in a QB net, it may be counterproductive to
constrain
$A$ to be real, since this constraint
may cause SEO decompositions of $A$ to be longer.


\bibitem{L-S}S.L. Lauritzen and D.J.
Spiegelhalter,
Jour. of the Royal Statistical Society {\bf B}
(1988) {\bf 50}, 157.


\end{thebibliography}
\end{document}